%% file: manuscript.tex
\documentclass[11pt]{article}
\usepackage[utf8]{inputenc}

\usepackage{amsmath}
\usepackage{amsfonts}
\usepackage{amssymb}
\usepackage{mathrsfs}

\usepackage{MnSymbol}

\usepackage[cal=boondox,scr=boondoxo]{mathalfa}

\usepackage{float}
\usepackage{subfig}
\usepackage{fancybox,graphicx}
\usepackage{subfig}
\usepackage{caption}
\usepackage{color}
\usepackage{authblk}

\usepackage[colorlinks]{hyperref}
\input{doiCmd} 

\usepackage{accents}
\usepackage[titletoc,title]{appendix}
\usepackage{cite}

\usepackage{mathtools}

\usepackage[top=2in, bottom=1.5in, left=1in, right=1in]{geometry}

\usepackage{tikz,xcolor}

\definecolor{lime}{HTML}{A6CE39}
\DeclareRobustCommand{\orcidicon}{%
	\begin{tikzpicture}
	\draw[lime, fill=lime] (0,0) 
	circle [radius=0.16] 
	node[white] {{\fontfamily{qag}\selectfont \scriptsize ID}};
	\draw[white, fill=white] (-0.0625,0.095) 
	circle [radius=0.007];
	\end{tikzpicture}
	\hspace{-2mm}
}

\foreach \x in {A, ..., Z}{%
	\expandafter\xdef\csname orcid\x\endcsname{\noexpand\href{https://orcid.org/\csname orcidauthor\x\endcsname}{\noexpand\orcidicon}}
}

\title{Comparative Analysis of Molecular Dynamics and Method of Moments in Two-Dimensional Concentric Circular Layers\footnote{This is the version of the article before peer review or editing, as submitted by an author to \textbf{Comparative Analysis of Molecular Dynamics and Method of Moments in Two-Dimensional Concentric Circular Layers} IOP Publishing Ltd is not responsible for any errors or omissions in this version of the manuscript or any version derived from it. The Version of Record is available online at 
\url{https://doi.org/10.1088/1361-648X/ad5baf} .}}

\author[1]{Robert Salazar\orcidA{}\thanks{\href{mailto:rp.salazar84@uniandes.edu.co}{rp.salazar84@uniandes.edu.co}}}
\author[1]{Cristian Cobos\thanks{\href{mailto:cristiand.coboss@ecci.edu.co}{cristiand.coboss@ecci.edu.co}} }
\author[3]{Diego Jaramillo\orcidC{}\thanks{\href{mailto:djaramillo86@uan.edu.co}{djaramillo86@uan.edu.co}}}
\author[2]{Camilo Bayona-Roa\orcidB{}\thanks{\href{mailto:cabayonar@unal.edu.co}{cabayonar@unal.edu.co}}}

\affil[1]{Direcci\'on de Ingenier\'ia Electr\'onica, Universidad ECCI, Bogot\'a, Colombia}
\affil[2]{Centro de Ingenier\'ia Avanzada Investigaci\'on y Desarrollo --- CIAID, Bogot\'a, Colombia}
\affil[3]{Departamento de Ciencias, Universidad Antonio Nari\~no, Bogot\'a, Colombia}

\begin{document}

    \maketitle

\begin{abstract}
In this manuscript, we undertake an examination of a classical plasma deployed on two finite co-planar surfaces: a circular region $\Omega_{in}$ into an annular region $\Omega_{out}$ with a gap in between. It is studied both from the point of view of statistical mechanics and the electrostatics of continua media. We employ a dual perspective: the first one is by using Molecular Dynamics (MD) simulations to find the system's positional correlation functions and velocity distributions. That by modeling the system as a classical two-dimensional Coulomb plasma of point-like charged particles $q_1$ and $q_2$ on the layers $\Omega_{in}$ and $\Omega_{out}$ respectively with no background density. The second one corresponds to a finite surface electrode composed of planar metallic layers displayed on the regions $\Omega_{in}$, $\Omega_{out}$ at constant voltages $V_{in}$, $V_{out}$ considering axial symmetry. The surface charge density is calculated by the Method of Moments (MoM) under the electrostatic approximation. Point-like and differential charges elements interact via a $1/r$ - electric potential in both cases. The thermodynamic averages of the number density, and electric potential due to the plasma depend on the coupling and the charge ratio $\xi=q_1/q_2$ once the geometry of the layers is fixed. On the other hand, the fields due to the SE depend on the layer's geometry and their voltage. In the document, is defined a protocol to properly compare the systems. We show that there are values of the coupling parameter, where the thermodynamic averages computed via MD agree with the results of MoM for attractive $\xi=-1$ and repulsive layers $\xi=1$. 
\end{abstract}

\section{Introduction}
\begin{figure}[H]
\centering 
\includegraphics[width=0.50\textwidth]{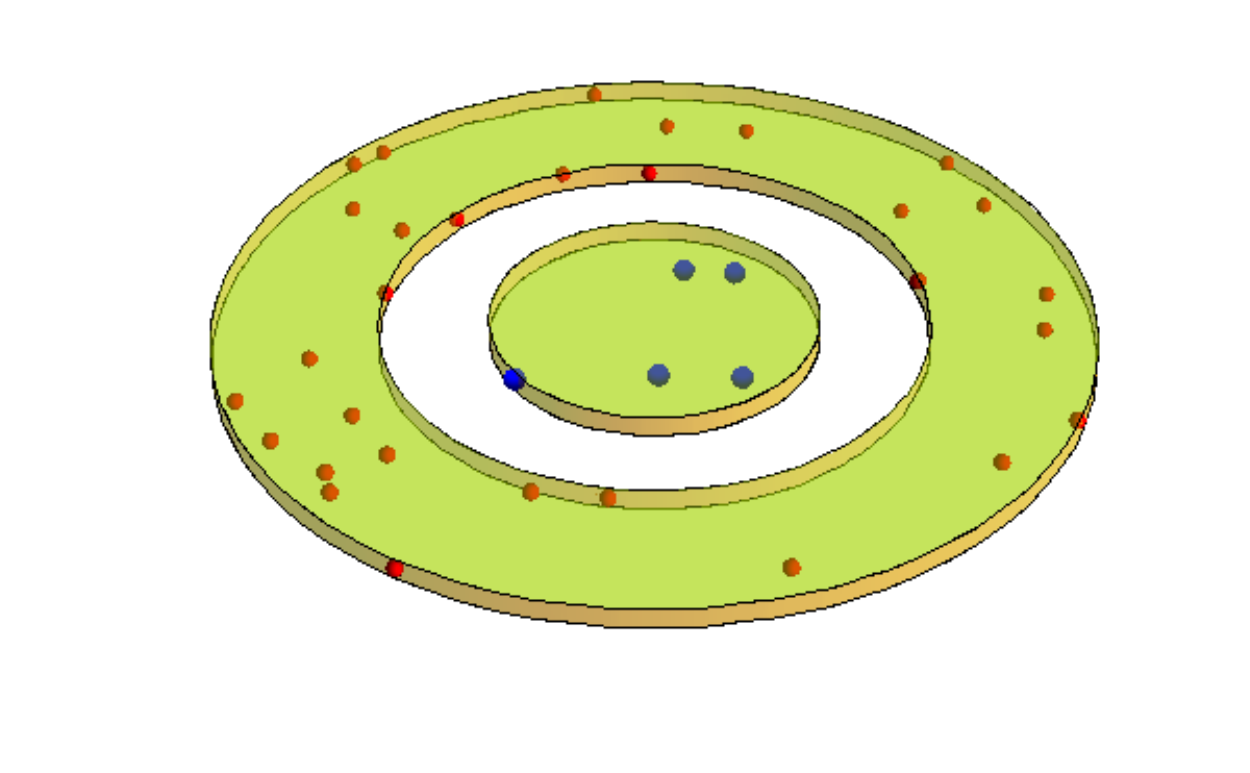}\hspace{0.5cm}
\includegraphics[width=0.3\textwidth]{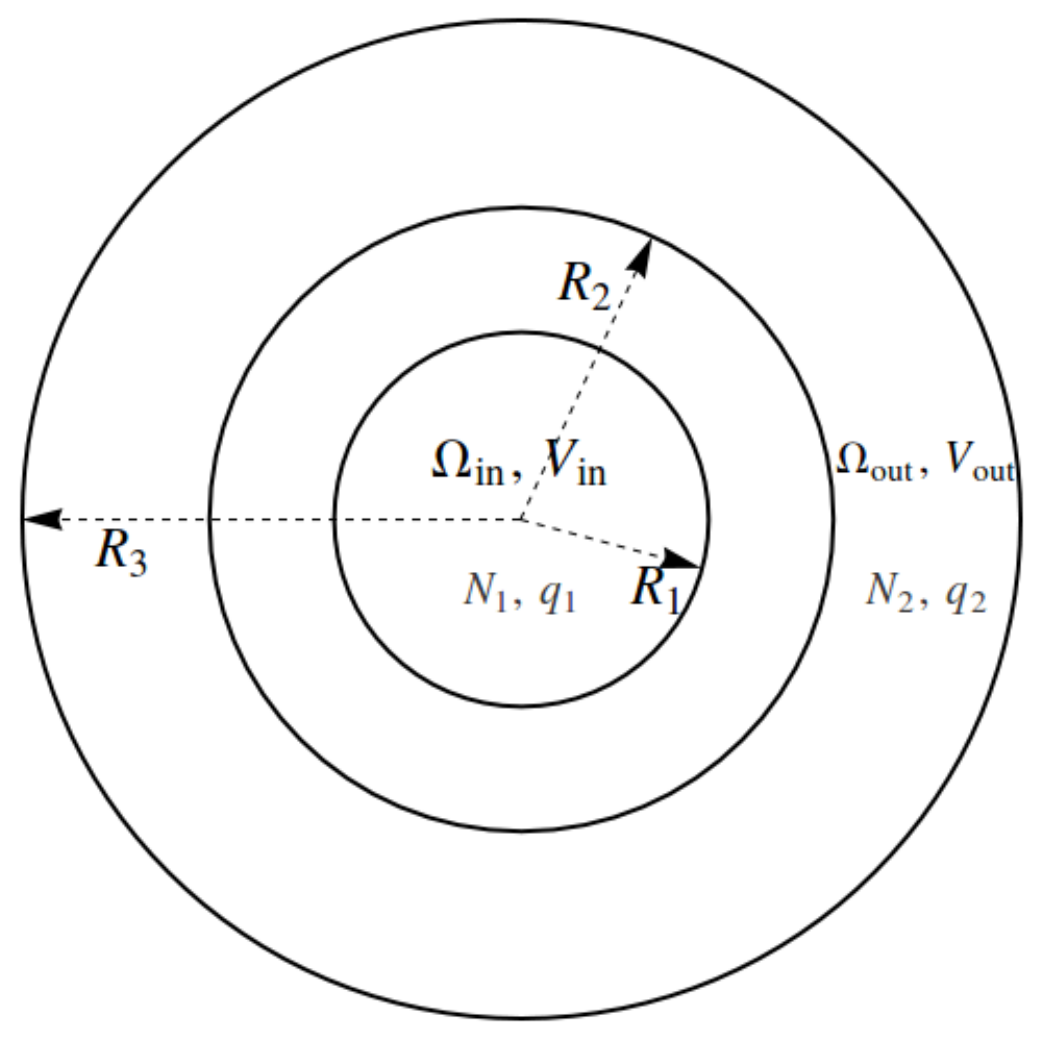}
    \caption[The System.]{The system. Two metallic sheets $\Omega_{in}$ and $\Omega_{out}$ with no background density charge having $N_1$ particles of charge $q_1$  and $N_2$ particles of charge $q_2$. }
\label{theSystemFig}
\end{figure}

In this study, we examine a metallic configuration comprising distinct layers denoted as $\Omega_{in}$ and $\Omega_{out}$, collectively forming a gapped Surface Electrode (SE) devoid of background charge density. The inner layer is a disk $\Omega_{in}= \left\{(x,y) : \sqrt{x^2+y^2} \leq R_1 \right\}\subset\mathbb{R}^2$ of radius $R_1=R$ and the outer layer is an annular region $\Omega_{out}= \left\{(x,y) :  R_2 \leq\sqrt{x^2+y^2} \leq R_3 \right\}\subset\mathbb{R}^2$. The system can be also modeled as two-dimensional plasma where within $\Omega_{in}$, reside $N_1$ particles with charge $q_1\in\mathbb{R}$, while $\Omega_{out}$ holds $N_2$ particles with charge $q_2\in\mathbb{R}$ (see Fig.~\ref{theSystemFig}).
The foregoing study draws inspiration from \cite{salazar2022electric}, where the gapped surface electrode was characterized by two-conductor flat regions with differing potentials and a distinct gap. Electric field and surface charge density analysis in that system involved resolving Laplace's equation with specified conditions.

The $R_1\longrightarrow R_2$ (gapless) and $R_3\longrightarrow \infty$ (semi-infinite outer layer) limits of the gapped SE were studied in \cite{romero2022monte}. The surface electrode was modeled as a two-dimensional classical Coulomb gas with $+q$ and $-q$ charges on the inner and outer electrodes. Equilibrium states and charge density were derived from Monte Carlo (MC) simulations. In that study, interactions followed an inverse power law potential $1/r$, where the coupling parameter $\Gamma$ has been found inversely proportional to the system's temperature and proportional to $q^2$. 

Furthermore, authors of Ref.~\cite{salazar2022electric} explored the electric vector potential and the Biot-Savart law analogy in electrostatics. These efforts set the zero stage for the present study, aiming to further analyze the system in question with the statistical mechanics approach. Subsequent sections detail our Monte Carlo methodology, results, and implications within the previous electrostatics descriptions of the system.

A strategy to study two-dimensional plasmas is by using Molecular Dynamics (MD) simulations \cite{dimonte2008molecular, zelener2018self, arkhipov2017direct, durniak2010molecular, zwicknagel1996molecular, calisti2011classical, scheiner2019testing, mithen2012molecular , CALISTI2009307}. In the present study, we shall perform simulations in the NVT-ensemble by using Nosé-Hoover and Langevin thermostats. Our goal is to study the thermodynamic averages of number density and the potential as a function of $\beta=k_B T$ for a fixed geometry of the SE. We aim to know if a coulomb plasma with a finite number of particles can be in agreement with the electrostatics of the SE.  

The motivation behind this study stems from the electrostatic behavior of the gapped surface electrode system comprising $\Omega_{in}$ and $\Omega_{out}$. While prior research has shed light on this system's electromagnetic fields, our work extends the investigation into the realm of statistical mechanics. By delving into the interplay between Molecular Dynamics simulations and the Method of Moments technique, we aim to uncover the underlying principles governing the system's surface charge density and its correlation with the Coulomb's interaction between charges.

This document is structured as follows: We first delve into the Method of Moments, focusing on the derivation of surface charge density for a single circular layer and exploring the distinctive electrostatic behavior originating from the interactions governed by a Coulomb potential. Then, we describe the Molecular Dynamics techniques, together with the experimental protocol that is followed to make comparisons. Subsequently, we present and analyze results obtained through both the MoM and MD methods. The statistical results are contrasted against analytic descriptions of SE. These findings are discussed within the context of electrostatics and statistical mechanics, highlighting their broader implications. Finally, we conclude by summarizing our contributions. 

\section{Method of Moments}
In this section, we delve into the main concepts of the Method of Moments and its application to our system. From the point of view of the electrostatics of continuum, (see Fig.~\ref{theSystemFig}-right) charge is continuously distributed over the SE layers. The potential of the layers is known from the problem definition, but the surface charge distribution $\sigma(\boldsymbol{r})$ is unknown. To calculate the surface charge distribution on the layers we use the MoM. That method is often employed to study classical electromagnetic radiation \cite{gibson2021method, li2004method, li1999method}, but it is also useful to derive surface charge densities in electrostatic systems \cite{harrington1987method, balanis2016antenna}. We start by examining the behavior of a single circular layer and exploring the derivation of surface charge density. Furthermore, we investigate the intricate interplay of interactions governed by the $1/r$ potential, shedding light on the unique electrostatic behavior of our system.

\subsection{Single circular layer}
First, we derive the surface charge density for a single circular layer configuration. This first step serves as the foundational study in a single component of the gapped SE system.
The integral formula of the electrostatic potential is given by
\[
\frac{1}{4\pi\epsilon_o} \int_{\Omega} \frac{\sigma(\boldsymbol{r}')}{|\boldsymbol{r}-\boldsymbol{r}'|}d^2\boldsymbol{r} = \Phi(\boldsymbol{r}\in\Omega) = V_o,
\]
with $\epsilon_o$ the permittivity of vacuum and $V_o$ the voltage of the circular layer $\Omega= \left\{(x,y) : \sqrt{x^2+y^2} \leq R \right\}\subset\mathbb{R}^2$ of radius $R$. Since the system has azimuthal symmetry then potential electrostatic potential in the space is independent of the angular $\phi$-coordinate $\Phi(\boldsymbol{r}) = \Phi(u,z)$. Therefore, the surface charge density is a function depending only of the radial $u=\sqrt{x^2+y^2}$ coordinate $\sigma(\boldsymbol{r}) = \sigma(u)$. Any location on the plate is given by $\boldsymbol{r}(u,\phi) = u\hat{u}(\phi)$, and the distance between two different points on the plate is
\[
|\boldsymbol{r}-\boldsymbol{r}'| = \sqrt{u^2+(u')^2 - 2uu' \hat{u}(\phi)\cdot\hat{u}(\phi')},
\]
with $\hat{u}(\phi) = (\cos\phi,\sin\phi,0)$ written in Cartesian coordinates. Hence, the potential on the metallic layer can be calculated from
\[
\frac{1}{4\pi\epsilon_o} \int_{0}^R\int_{0}^{2\pi} \frac{\sigma(u)}{ \sqrt{u^2+(u')^2 - 2uu'\cos(\phi-\phi')}}u'du'd\phi' = \Phi(\boldsymbol{r}\in\Omega) = V_o.
\]
We aim to obtain the surface charge density $\sigma$ from the previous integral equation. The strategy is to expand $\sigma$ into a truncated series of functions as follows,
\[
\sigma(u) = \sum_{n=0}^N \sigma_n f_n(u),
\]
with $\left\{f_n(u)\right\}_{n=1,\ldots,N}$ the set of basis functions where $\sigma_n$ is the $n$-th coefficient for the surface charge density in that basis. Then, 
\[
\frac{1}{4\pi\epsilon_o} \sum_{n=0}^N \sigma_n \int_{0}^R\int_{-\phi}^{2\pi-\phi}  \frac{f_n(u')}{ \sqrt{u^2+(u')^2 - 2uu'\cos(\beta)}}u'du'd\beta' = \Phi\left((u,\phi,0)\in\Omega\right) = V_o,
\]
where we have used the change of variable $\beta = \phi'-\phi$. Note that the choice of $\phi\in[0,2\pi]$ only ranges the surface of the plate, then
\[
\frac{1}{4\pi\epsilon_o} \sum_{n=0}^N \sigma_n \int_{0}^R \left[\int_{0}^{2\pi}  \frac{d\beta}{ \sqrt{u^2+(u')^2 - 2uu'\cos(\beta)}}\right]f_n(u')u'du' = \Phi\left((u,0,0)\in\Omega\right) = V_o.
\]
The angular integral can be evaluated straightforwardly since it is related to the complete elliptic function of the first kind\footnote{This integral can be simplified by introducing
\[
\zeta(u',u) := \sqrt{(u')^2+u^2 - 2u'u}=|u-u'|,\hspace{0.5cm}\mbox{and}\hspace{0.5cm} \xi := \frac{4u'u}{\zeta(u',u)^2}=\frac{4u'u}{|u-u'|^2}.
\]
Changing the $\beta$ variable with $\beta=2t$ leads to
\begin{align}
    (u')^2+u^2-2u'u\cos\beta &= (u')^2+u^2-2u'u + 2 u'u(1-\cos\beta) \nonumber\\
                            &= ((u')^2+u^2-2u'u) + 4 u'u\sin^2t \nonumber\\
                                   &= \zeta(u',u)^2 + \xi\zeta(u',u)^2 \sin^2t \nonumber\\  
                                   &= \zeta(u',u)^2(1+\xi\sin^2t)=|u-u'|^2(1+\xi\sin^2t).   \nonumber
\end{align}
Hence, the integrand can be written as
\[
\int_{0}^{2\pi}  \frac{d\beta}{ \sqrt{u^2+u'^2 - 2uu'\cos(\beta)}} = \frac{4}{|u-u'|}\int_{0}^{\pi/2} \frac{dt}{\sqrt{(1+\xi\sin^2t)}} =  \frac{4}{|u-u'|}K\left( -\xi\right).
\]},
\begin{equation}
\int_{0}^{2\pi}  \frac{d\beta}{ \sqrt{u^2+(u')^2 - 2uu'\cos(\beta)}} = \frac{4}{|u-u'|}K\left(-\frac{4uu'}{|u-u'|^2}\right),
\label{auxAngularEq}
\end{equation}
where the elliptic function is given by
\[
K(\chi) = \int_{0}^{\pi/2} \frac{d\theta}{\sqrt{1-\chi\sin^2\theta}}.
\]

\begin{figure}[H]
\centering 
\includegraphics[width=0.55\textwidth]{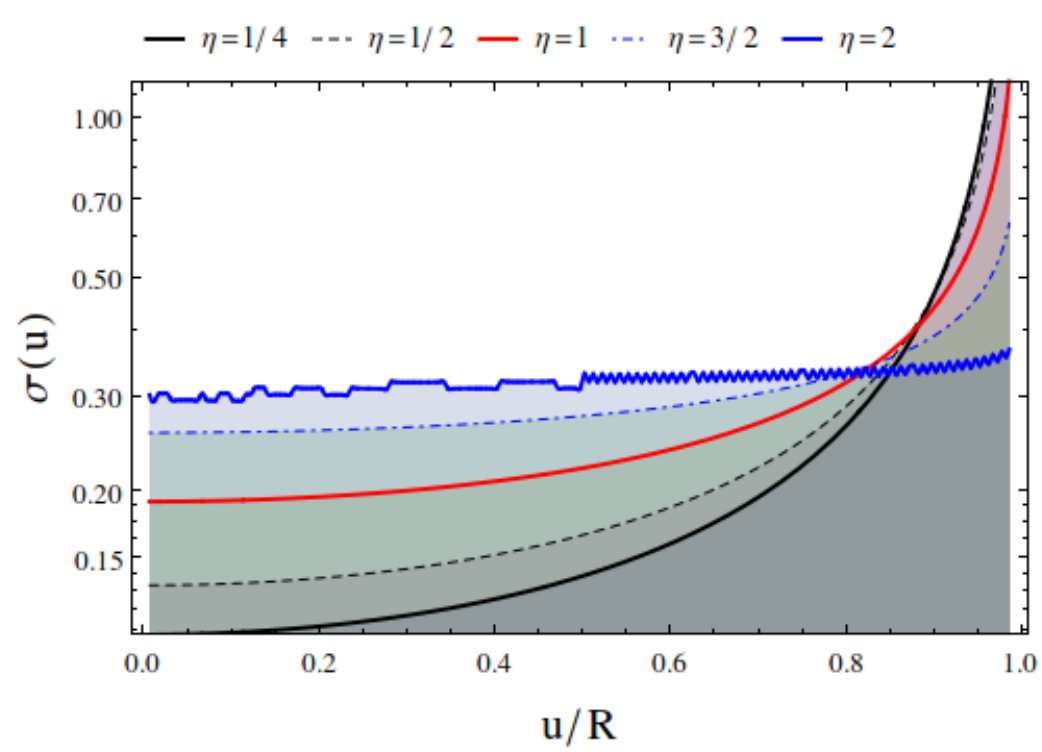}
\includegraphics[width=0.44\textwidth]{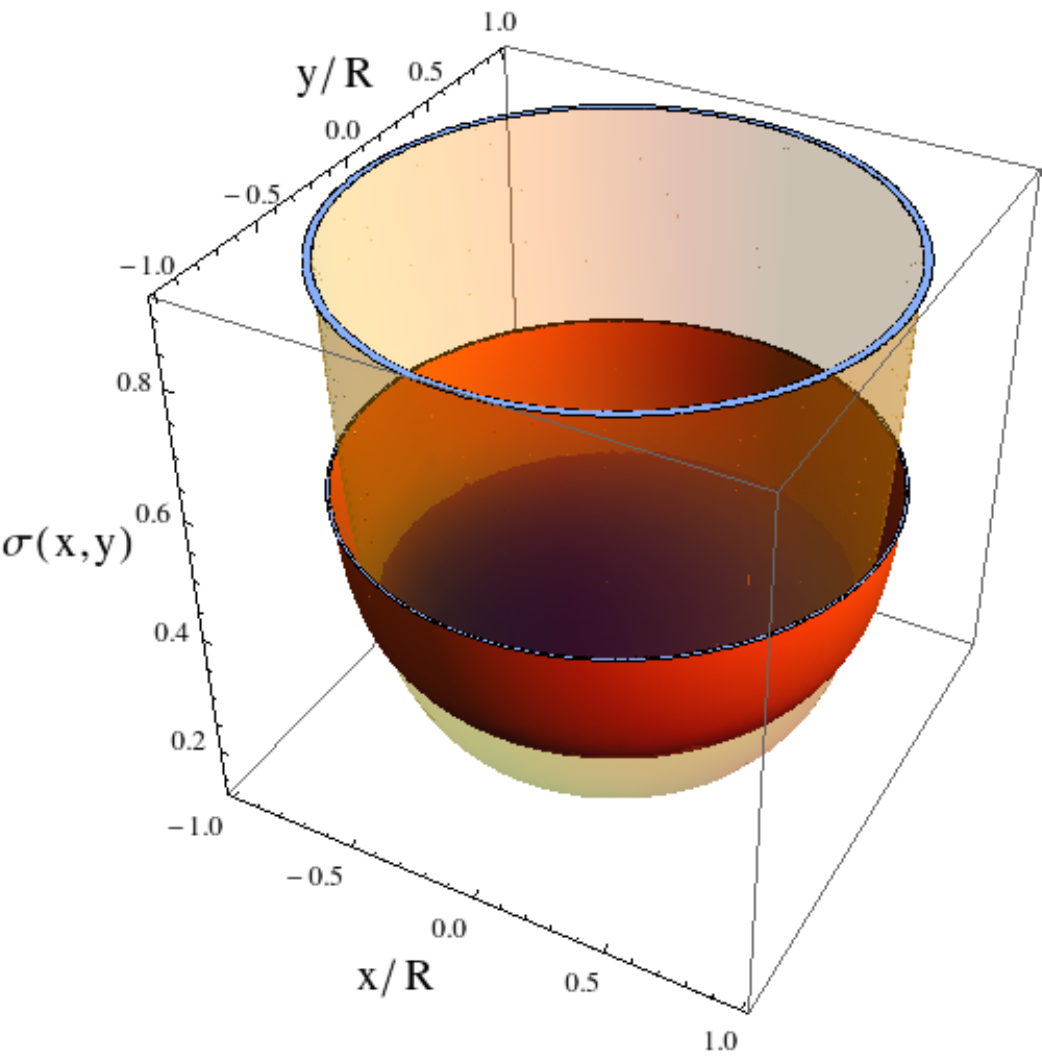}
    \caption[The curve.]{Surface charge density of the singular circular plate at a constant voltage according to MoM. (left) Profile of the charge density.  (right) Surface density charge for $\eta=1/2$ (transparent surface) and $\eta=3/2$ (solid surface).}
\label{sigmaFig}
\end{figure} 

The potential evaluated on the sheet is computed from
\[
\frac{1}{\pi\epsilon_o} \sum_{n=0}^N \sigma_n \int_{0}^R \frac{f_n(u')}{|u-u'|}K\left(-\frac{4uu'}{|u-u'|^2}\right)u'du' = \Phi\left((u,0,0)\in\Omega\right) = V_o.
\]
In this case, we can choose the value of $u$ in the interval $[0, R]$. We shall define $\mathcal{N}+1$ equally spaced discrete values at the radial coordinate $u_0,\ldots,u_\mathcal{N}$ with $u_n = n R / \mathcal{N}$, to define $\mathcal{N}$ intervals $U_n := [u_n, u_{n+1}]$ with centers located at $u_n^{(c)} = (u_n + u_{n+1})/2$, then 
\[
\frac{1}{\pi\epsilon_o} \sum_{n=0}^N \sigma_n \int_{0}^R \frac{f_n(u')}{|u_m^{(c)}-u'|}K\left(-\frac{4u_m^{(c)}u'}{|u_m^{(c)}-u'|^2}\right)u'du' = \Phi\left((u_m^{(c)},0,0)\in\Omega\right) = V_o,
\]
which can be written as a linear set of algebraic equations that can be inverted to find the $\sigma_n$ coefficients, as follows
\[
\sum_{n=0}^\mathcal{N} M_{mn}\sigma_n = \Phi_m  \hspace{0.5cm}\therefore\hspace{0.5cm} \sigma_m=\sum_{n=0}^\mathcal{N} M_{mn}^{-1} \Phi_n = \sum_{n=0}^\mathcal{N} M_{mn}^{-1}\{V_o\}_n, 
\]
with
\[
M_{mn} = \frac{1}{\pi\epsilon_o}  \int_{0}^R \frac{f_n(u')}{|u_m^{(c)}-u'|}K\left(-\frac{4u_m^{(c)}u'}{|u_m^{(c)}-u'|^2}\right)u'du' 
\]
being a square matrix and $\Phi_m = \Phi\left((u_m^{(c)},0,0)\in\Omega\right)$ a vector of constant entries $V_o$.

There are several choice of basis functions $\{f_u(n)\}_{n=0,\ldots,\mathcal{N}}$. One of the simplest choices is the $C_0$ basis defined as
\[
f_n(u) = \begin{cases}
  1  & u \in U_n=[u_n, u_{n+1}], \\
  0 & \mbox{otherwise},
\end{cases}
\]
for which the $M_{mn}$ matrix becomes
\[
M_{mn} = \frac{1}{\pi\epsilon_o}  \int_{U_n} \frac{u'du'}{|u_m^{(c)}-u'|}K\left(-\frac{4u_m^{(c)}u'}{|u_m^{(c)}-u'|^2}\right). 
\]
If the number of the interval divisions is large $\mathcal{N}>>1$ one can use a mid-point approximation, leading to the simpler discrete term
\[
M_{mn} = \frac{1}{\pi\epsilon_o}  \frac{u_n^{(c)}\Delta u_n}{|u_m^{(c)}-u_n^{(c)}|}K\left(-\frac{4u_m^{(c)}u_n^{(c)}}{|u_m^{(c)}-u_n^{(c)}|^2}\right)    \hspace{1.0cm}\mbox{for}\hspace{1.0cm} m \neq n, 
\]
with $\Delta u_n=u_{n+1}-u_n$. However, it is often advisable to perform numerical integration through sophisticated techniques to obtain accurate values of $M_{mn}$. In this manuscript, these integrals are evaluated by using the \textit{Global Adaptive} method set by default in function \textit{NIntegrate} of Wolfram Mathematica 9.0 \cite{wolfram2012version}. The MoM is a versatile method and it can be easily generalized for other long-range potentials of the form $1/r^{\eta}$ with $\eta\in\mathbb{R}^+$ (as demonstrated in Appendix~\ref{generalizationAppendixLbl}). For the case of a single circular plate, the voltage $V_o$ plays a scaling role for the charge density, thus one can set that voltage as one. In Fig.~\ref{sigmaFig}-left we show the surface charge density profile for standard $1/r$-interaction potential (red-solid line) for the circular plate. It occurs that surface charge density in the electrostatic approximation diverges at the sheet border, that is $\lim_{u\to R}\sigma(u) = \infty$. Numerically, we do not reach that limit since we compute the vector $\sigma(u_m^{(c)}) = \sigma_m$ and the largest value $\sigma_N$ is located at $R-\Delta u_\mathcal{N}/2$ which approaches to the border as $\mathcal{N}\longrightarrow\infty$, where $\sigma_{\mathcal{N}\to\infty} \longrightarrow \infty$. 

\subsection{Two layers}
The MoM for concentric circular layers (see Fig.~\ref{theSystemFig}) can be straightforwardly generalized by following an analogous procedure to the one described for one circular plate. A generalization of this procedure to the gapped SE is presented in Appendix~\ref{appMoMTwoLayersSec}. Summarizing, two basis functions $\{f_n(u)\}_{n=1,\ldots,2\mathcal{N}}$ are employed to expand the surface charge density: the first $n=1,\ldots,\mathcal{N}$ bases are employed for the inner layer and the complementary  $n=\mathcal{N}+1,\ldots,\mathcal{2N}$ are used for the second layer. The surface charge density between these elements is  
\begin{equation}
\sigma_m = V_{in} \sum_{n=1}^{\mathcal{N}}   M_{m,n}^{-1} + V_{out} \sum_{n=\mathcal{N}+1}^{2\mathcal{N}}   M_{m,n}^{-1}  \hspace{1.0cm}m=1,\ldots,2\mathcal{N},
\label{MoMSigmaTwoSheetsEq}
\end{equation}
with $M_{m,n}$ a $2\mathcal{N} \times 2\mathcal{N}$ matrix given by
\[
M_{mn}=
    \int_{u_{n-1}}^{u_{n}} I(u_m^{(c)},u') u'du'  \hspace{0.25cm}\mbox{\textbf{if}}\hspace{0.25cm} n=1,\ldots,\mathcal{N},     \hspace{0.25cm}\mbox{\textbf{otherwise}}\hspace{0.25cm}\int_{u_{n}}^{u_{n+1}} I(u_m^{(c)},u') u'du'  ,
\]
and
\[
I(u,u') = \frac{1}{\pi\epsilon_o}\frac{1}{|u-u'|}K\left(-\frac{4uu'}{|u-u'|^2}\right).
\]



\section{Molecular Dynamics}
In this section, the SE system is modeled as $N_1$, $N_2$ distribution of classical electric charges (with nominal values $q_1$ and $q_2$) confined to planar geometries $\Omega_1$ and $\Omega_2$, respectively (see Fig.~\ref{theSystemFig}). The coarse-grained method is used to model charged particles, which collide elastically with the layer boundaries $\partial \Omega_1$,  $\partial \Omega_2$ and interacting with each other via a Coulomb-like potential of the form  
\[
\Phi(\boldsymbol{r}_{ij}) = \frac{1}{4\pi\epsilon_o} \frac{q_i q_j}{|\boldsymbol{r}_i - \boldsymbol{r}_j|}  \hspace{0.5cm}\mbox{if}\hspace{0.5cm} |\boldsymbol{r}_i - \boldsymbol{r}_j| > 2b, \hspace{0.5cm}\mbox{otherwise}\hspace{0.5cm}\infty.
\]

We employ two different MD techniques. In the first one, we consider hard-disk interactions with a radius of particles denoted by $b$, in addition to Coulomb interactions. For that technique, we employ the Langevin thermostat. In the second simulation, we eliminate hard-disk interactions and employ the Nose-Hoover thermostat. In both cases, we implement a Verlet algorithm to numerically integrate the equations of motion. Point-like charges remain on the layers, ensuring that particles in $\Omega_1$ did not mix with particles in $\Omega_2$.

For convenience, we introduce the parameter $\xi = q_2/q_1$, i.e., the ratio between the two types of charges. We restrict that parameter in the range $-1 \leq \xi \leq 1$ fixing the charge of the inner layer $q_1 = q$ and varying the charge of the outer layer in the range $-q \leq q_2 \leq q$. 

We do not compensate the particles' repulsion on the layers e.g. by introducing a neutralizing background on each layer, since we look for charge configurations that make layers equipotential surfaces, as it occurs in electrostatics. In every simulation, we set the number of particles to $N_1=N_2=N/2$ on the layers with $N\in 2\mathbb{N}$ the total number of particles, thus being a globally neutral system when $\xi=-1$. 

In MD simulations, we compute the number density  $n(\boldsymbol{r})$ (i.e number of particles per unit of area) that enables the computation of the surface charge density, as follows,
\begin{equation}
\sigma(u) =  \begin{cases}
    n(u) q &  0 \leq u \leq  R_1,  \\    
    n(u) \xi q & R_2 \leq u \leq  R_3, \\
    0 & \mbox{otherwise}.
\end{cases}
    \label{surfaceChargeAuxEq}
\end{equation}
The number density is computed by counting the particles in 
$$\boldsymbol{ann}(u, \Delta u) := \left\{ (u',\phi) \in \mathbb{R}^2 : u < |(u',\phi)| < u + \Delta u \right\}$$ e.g. concentric annular regions of thickness $\Delta u$ and dividing by the area of those regions.

\section{Numerical experiment protocol}
\label{MethologySectionLbl}
In order to perform proper comparisons between the results of the approaches in the study, we follow an experimental design. This is necessary because the starting parameters of MoM and MD are different, even for the common geometry defined by $R_1$, $R_2$ and $R_3$. Regarding the MoM, one has to set the layers' potential $V_{in}$ and $V_{out}$. On the other hand, MD requires establishing the number of particles of each layer $N_1$, $N_2$, the charge ratio $\xi$, and the temperature $T$, since the system is in the canonical ensemble. Of course, the starting parameters of MoM and MD are related by the electrostatic energy
\[
U_e = \int_{\Omega_{in}\cup\Omega_{out}} \Phi(\boldsymbol{r}) \sigma(\boldsymbol{r}) d^2\boldsymbol{r} = V_{in} \int_{\Omega_{in}} \sigma(\boldsymbol{r}) d^2\boldsymbol{r} + V_{out} \int_{\Omega_{out}} \sigma(\boldsymbol{r}) d^2\boldsymbol{r} = V_{in} Q_{in} + V_{out} Q_{out}, 
\]
with $Q_{in} = N_1 q_1$ and $Q_{out} = N_2 q_2$ the total charge of each layer. This is the energy stored, or conversely, the the amount of energy spent to charge both layers. That energy is always positive since
\[
U_e = \frac{1}{2} \epsilon_o \int_{\mathbb{R}^3} \boldsymbol{E}^2(\boldsymbol{r}) d^3\boldsymbol{r} \geq 0,
\]
with $\boldsymbol{E}(\boldsymbol{r})$ the electric field on the $\mathbb{R}^3$-space due to the surface charge distribution. It implies that
\[
V_{in} N_1 q_1 + V_{out} N_2 q_2 \geq 0,  
\]
thus the potential employed in the MoM and the total charge on the layer set in MD are conditioned by the electrostatic energy. In order to avoid the use of scaling factors in the computation of surface charge density comparisons we proceed as follows:

\begin{itemize}
    \item Step 1. We perform a MD simulation of the system by setting the number of particles $N = 2N_1 = 2N_2$, the charge ratio $\xi$, and the thermostat target temperature.  

    \item Step 2. Once the system is in equilibrium, the particles' density $\langle n(u) \rangle $ is computed from the thermodynamic average of configurations. The surface charge density is calculated using Eq.~(\ref{surfaceChargeAuxEq}).

    \item Step 3. The average potential of those several configurations is computed from
    \begin{equation}
    \langle \Phi(u,z) \rangle = \frac{1}{4\pi\epsilon_o} \sum_{n} \langle \sigma(u_n) \rangle u_n \delta u_n \frac{4}{\sqrt{(u-u_n)^2+z^2}}K\left(-\frac{4uu_n}{(u-u_n)^2+z^2}\right),
    \label{thermodynamicAvergaPhiEq}
    \end{equation}
    with $u_n$'s radial positions on the layers. Brackets $\langle \cdot \rangle$ indicate thermodynamic average on equilibrated configurations. 

    \item Step 4. The potential of the layers is estimated as follows
    \[
    V_{in} = \frac{1}{R_1} \int_{0}^{R_1}  \langle \Phi(u, \delta ) \rangle du\hspace{0.25cm}\mbox{and}\hspace{0.25cm}V_{out} = \frac{1}{R_3-R_2} \int_{R_2}^{R_3}  \langle \Phi(u, \delta ) \rangle du,
    \]
    with small value of $\delta \approx 0$. These integrals are numerically approximated.  
    \item Step 5. The MoM is fed with the potential values $V_{in}$ and $V_{out}$ from the previous step.

\end{itemize}

\section{Results}
In this section, we present the comparisons between MD and MoM simulations. We perform MD simulations in the NVT-ensemble by using Nosé-Hoover and Langevin Thermostats \cite{kumar2022introduction,leimkuhler2015molecular}, by placing $N_1=500$ particles in the inner layer and $N_2=500$ in the outer one, varying the charge ratio $\xi$ and the parameter $\beta=k_B T$. Typical runs of MD simulations consist of $5\times10^6$ steps with a time step of $\Delta t=0.001\tau$. The number of MD steps to reach equilibrium are in the order of $1.5\times10^6$. 

For each set of parameters e.g. $\xi$, $\beta$ and fixed geometry, we feed the MoM with the thermodynamic average scalar potential near the layers calculated via MD. The MoM is run with 50 basis functions for each layer.  

In Subsection~\ref{subsectionCuplingEffectEq}, the coupling effect on the system behavior is discussed. In general, a finite N-system which is far from the thermodynamic limit cannot be described by the classical and continuum electrostatics in the $\beta \to \infty$ (weak coupling) and $\beta \to 0$ (strong coupling) limits. However, it is possible to find finite values of $\beta$ where the system behaves according to continuum electrostatics. Subsection~\ref{gaplessLimitLbl} is devoted to checking the MD simulations and the MoM calculations of surface density with the exact electrostatic result in the SE gapless limit. The last subsection compares MD and MoM results in the SE gapped case.

\subsection{The coupling effect over the finite SE system}
\label{subsectionCuplingEffectEq}
\begin{figure}[H]
\centering 

\begin{minipage}[b]{0.45\linewidth}
\centering 
\includegraphics[width=0.95\textwidth]{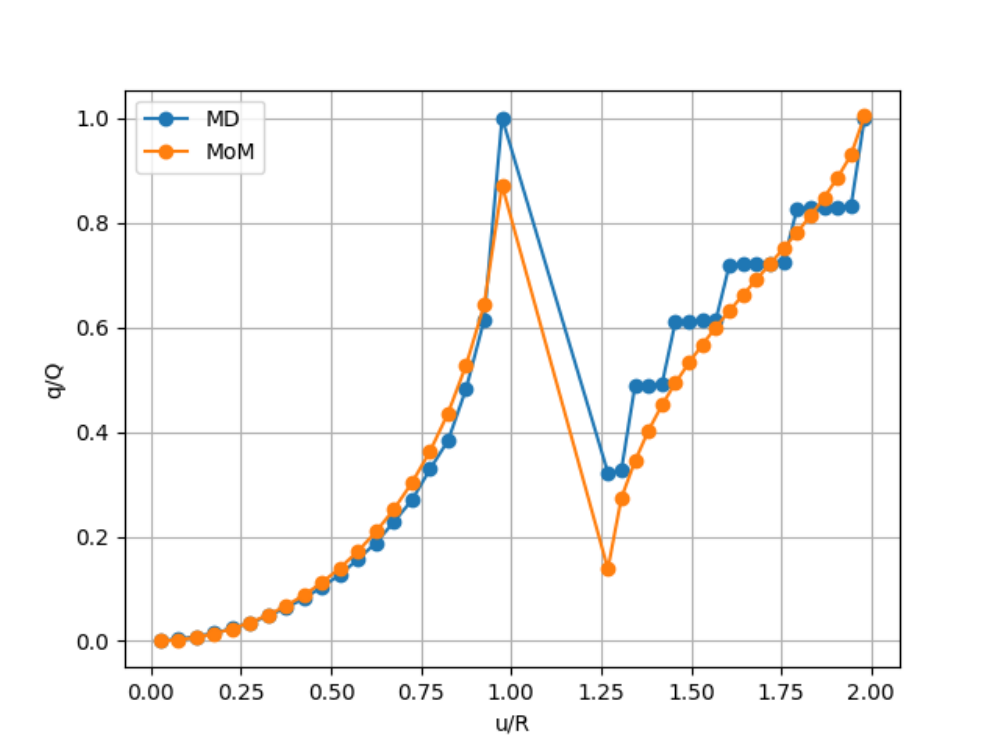}
\caption*{(a) Integrated charge via MD and MoM}
\end{minipage}
\begin{minipage}[b]{0.45\linewidth}
\centering 
\includegraphics[width=0.95\textwidth]{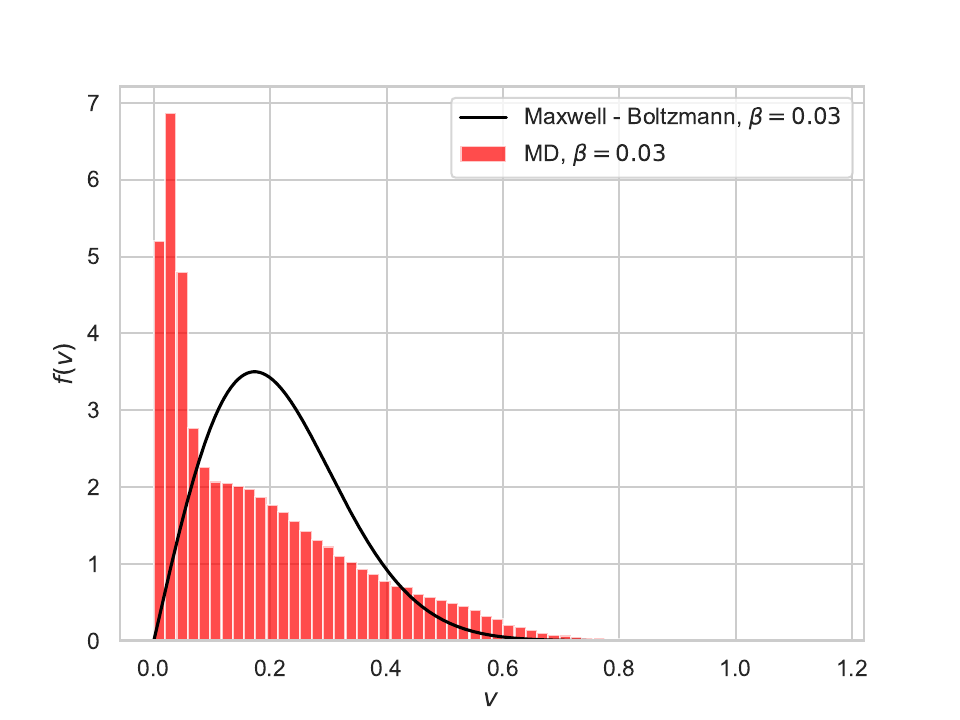}
\caption*{(b) Speed distribution}
\end{minipage}
\begin{minipage}[b]{0.45\linewidth}
\centering 
\includegraphics[width=0.95\textwidth]{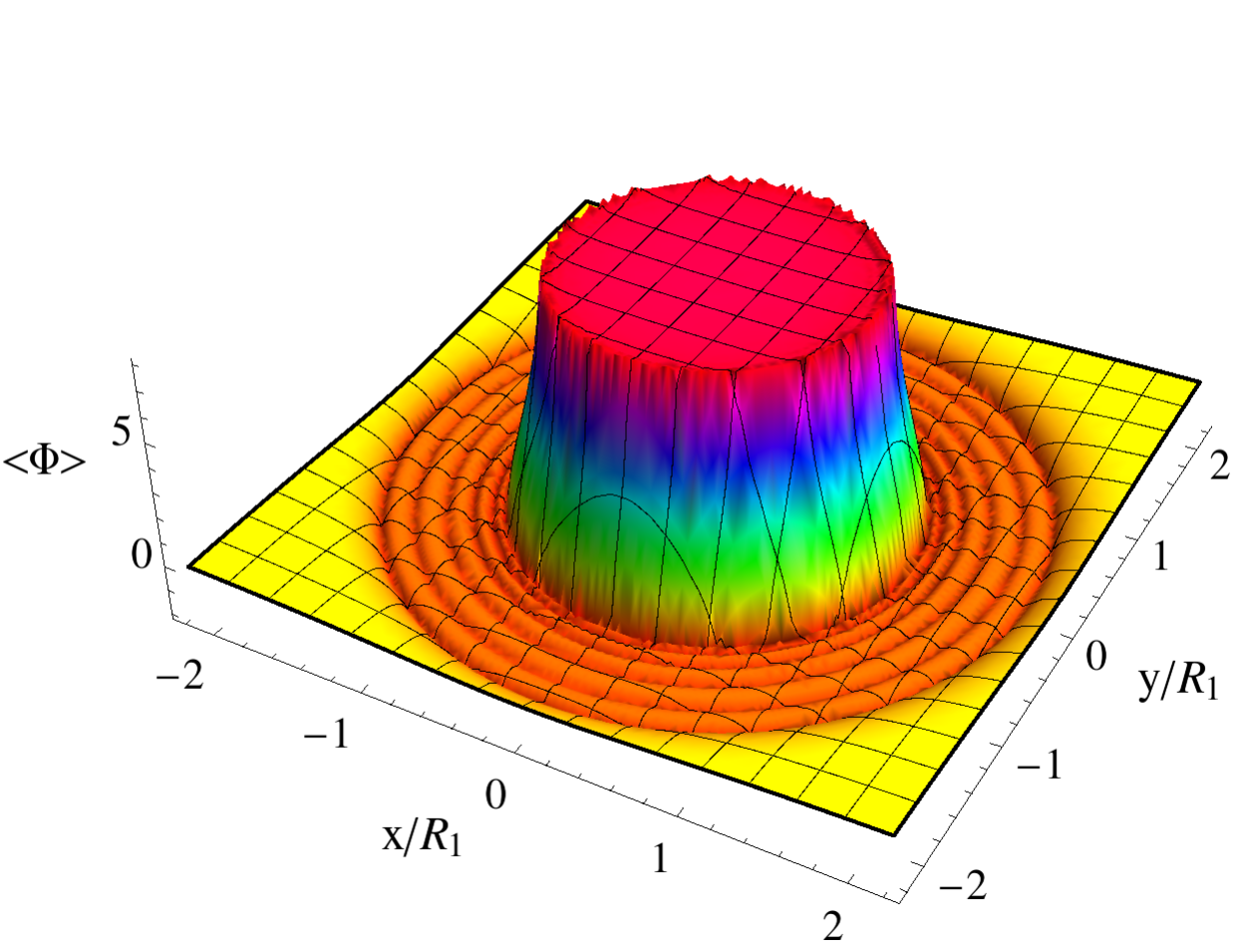}
\caption*{(c) Potential at $z=z_1=R_1/100$.}
\end{minipage}
\begin{minipage}[b]{0.45\linewidth}
\centering 
\includegraphics[width=0.95\textwidth]{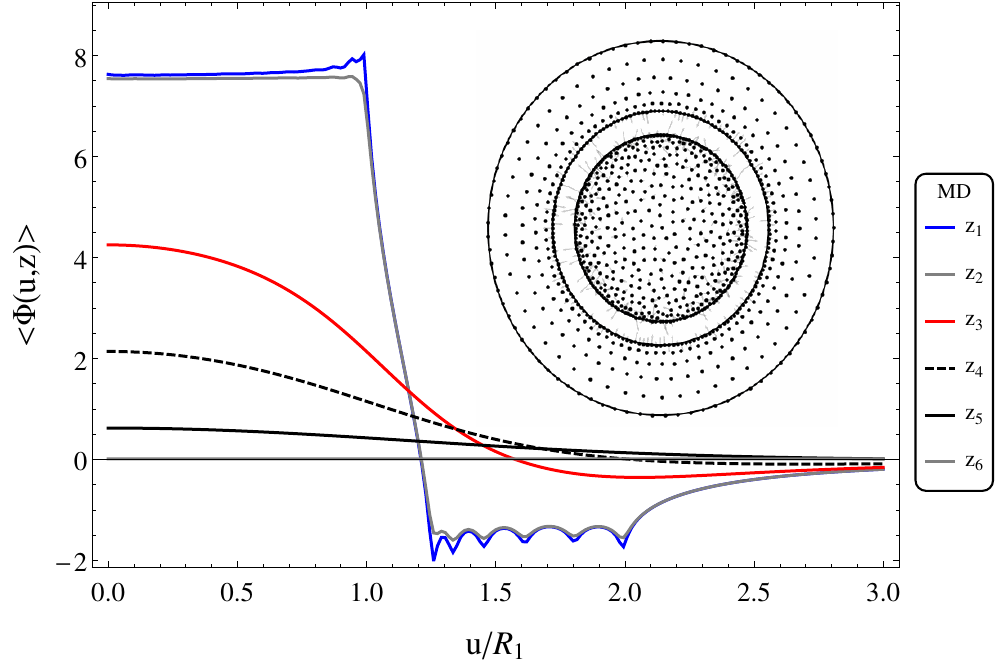}
\caption*{(d) Averaged Potential}
\end{minipage}
    \caption[Total charge.]{Strong coupling behaviour for $\xi=-1$ and $\beta=0.03$ MD results. The MD results corresponds to the Nosé-Hoover thermostat. The values of $z$ for (d) are $z_1=R_1/100$, $z_2=R_1/50$, $z_3=R_1/2$, $z_4=R_1$,$z_5=2R_1$, and $z_6=10R_1$.  }
\label{StrongCouplingFig}
\end{figure}

In general, the MD thermodynamic averages and the electrostatic predictions may diverge far from the thermodynamic limit $N \to \infty$, depending on the coupling between particles. For finite-size systems, the coupling between particles can change drastically the particle density distribution on the sheets, and the behavior of the system cannot be simply described by standard electrostatics of continuous media.  To exemplify this situation, the behavior of the finite system in a strong coupling regime is considered. Fig.~\ref{StrongCouplingFig} shows the results of a MD simulation with a Nosé-Hoover thermostat at $\beta=k_B T=0.03$. The parameters of the system are set as $N_1=N_2=500$, $R_1=4$, $R_2=5 R_1/4$, and $R_3=2R_1$, with opposite change $\xi=-1$. At this value of $\beta$, the electric interaction between particles is large enough in comparison with their kinetic energy, such that it \textit{freezes} the system. 

Technically, a non-crystalline solid is obtained at the $\beta \rightarrow 0$ limit via MD. At $\beta = 0.03$ the speed distribution $f(v)$ found through MD does not match the speed distribution of the classical two-dimensional ideal system:  
\begin{equation}
f(v) = \frac{mv}{\beta} \exp(-\frac{1}{2}\beta m v^2)  \hspace{0.5cm} \mbox{Maxwell speed distribution of a 2D ideal gas}
     \label{MaxwellBoltzmannDistribuitonEq}
\end{equation}
at the same $\beta$, since the histogram shows that there are fixed particles (Fig.~\ref{StrongCouplingFig}-b) due to the coupling. 

At the strong coupling regime $\beta \rightarrow 0$, the particles at the outer layer are arranged in concentric rings, as shown in the inset plot of Fig.~\ref{StrongCouplingFig}-d. On the other hand, the particles in the inner layer do not generate ring arrangements at $\beta=0.03$.  

Fig.~\ref{StrongCouplingFig}-c shows the normalized integrated charge $q/Q$ on each layer given by
\begin{equation}
   q(u) = 2\pi\begin{cases}
     \int_{0}^u \sigma(u') du' & \mbox{inner sheet,} \\
     \int_{R_2}^u \sigma(u') du' & \mbox{outer sheet,}
\end{cases} 
\label{integratedChargeIIEq}
\end{equation}
with $Q=q(R_1)$ and $Q=q(R_3)$ the total charge of each layer.

One can observe that the integrated charge via MD and MoM are in agreement in the inner layer, but they are not for the outer one as it is shown in Fig.~\ref{StrongCouplingFig}-a. The plot of $q/Q$ at the outer layer via MD has a stairs-like profile. Those jumps on $q(u)$ for $u \in [R_2,R_3]$ are due the circular arrangements of the particles at the outer layer. These circular arrangements of particles also affect the thermodynamic average potential near the outer layer, as it is shown in Fig.~\ref{StrongCouplingFig}-c.

\begin{figure}[H]
\centering 
\includegraphics[width=0.32\textwidth]{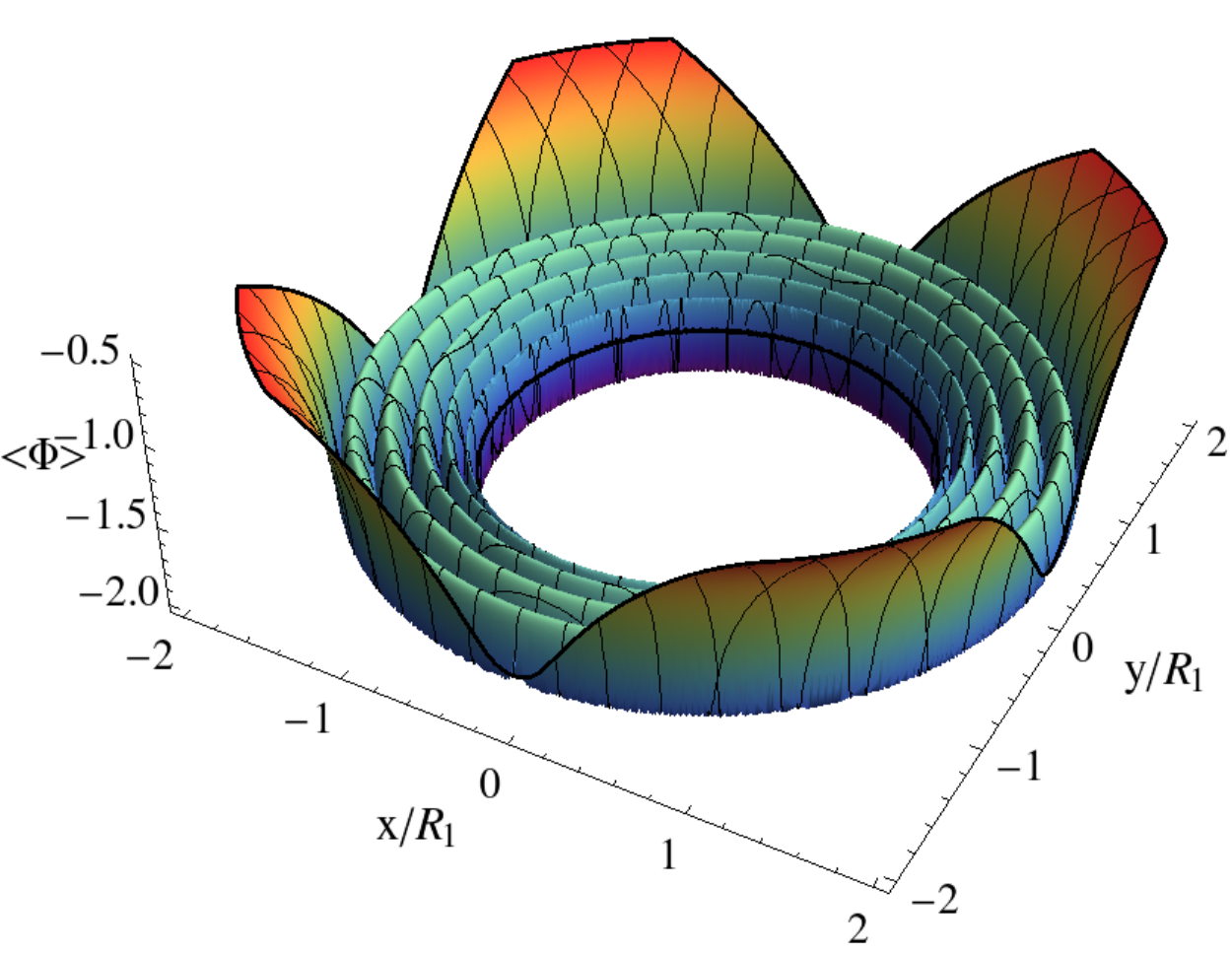}
\includegraphics[width=0.32\textwidth]{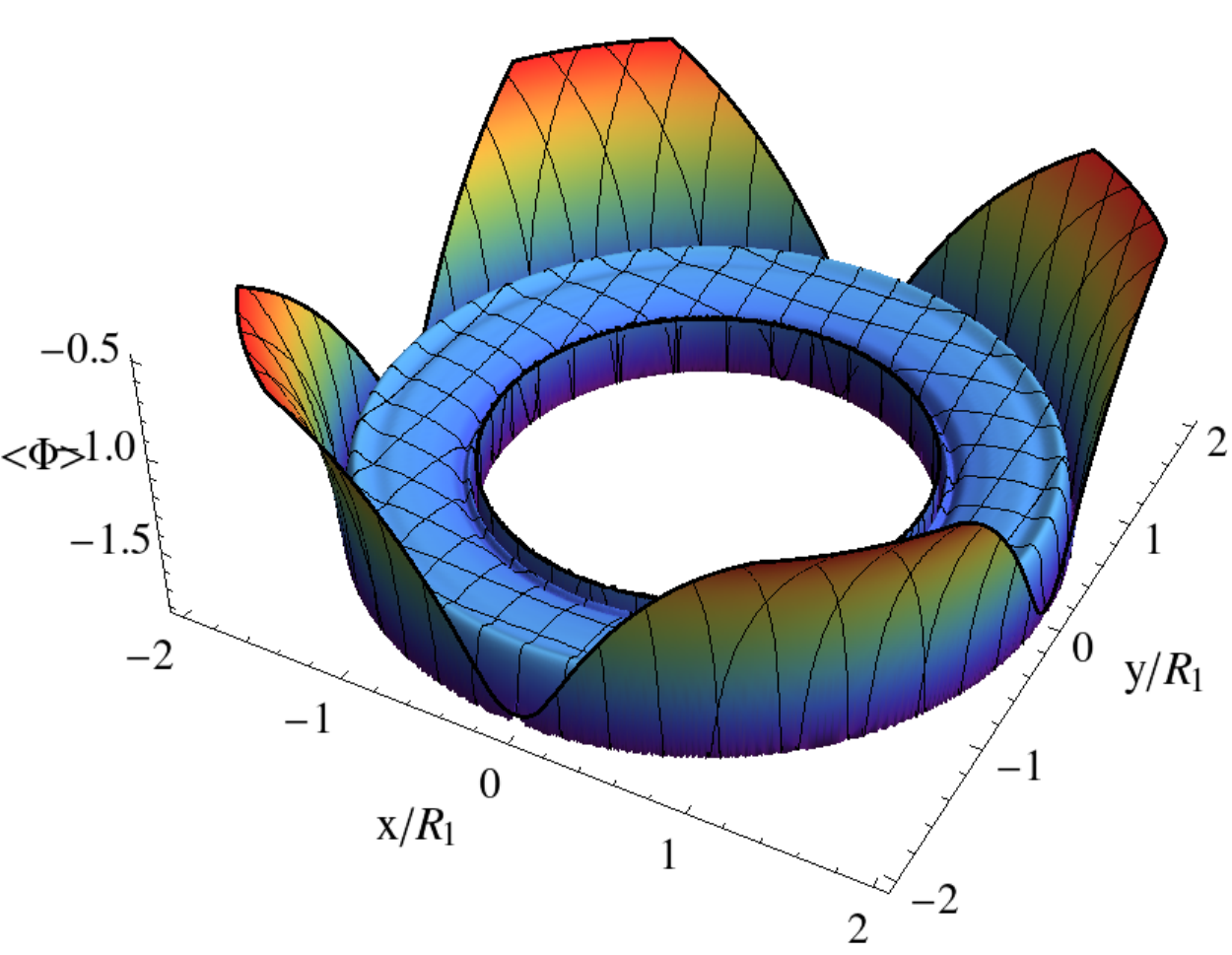}
\includegraphics[width=0.32\textwidth]{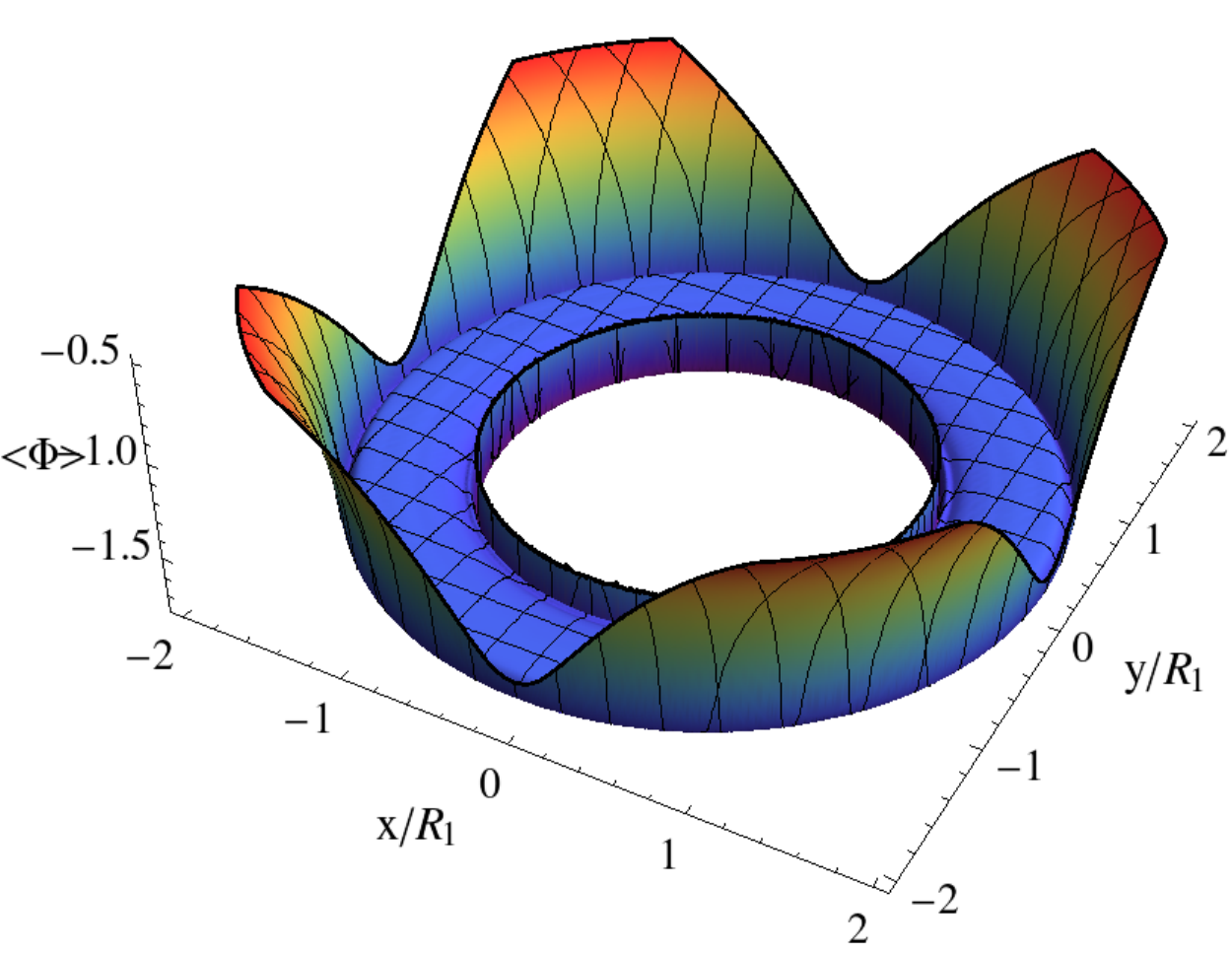}
    \caption[Total charge.]{MD average potential near the outer layer at the plane $z=z_1=R_1/100$. Molecular dynamics simulations with $\beta=0.03, 0.06,$ and $0.09$ (left to right).  }
\label{betaBehaviourOuterLayerFig}
\end{figure}

The thermodynamic average potential $\langle \Phi(u,z) \rangle$ computed via MD is shown in Fig.~\ref{StrongCouplingFig}-d. At $z=0.004$ and $z=0.04$, the average potential shows peaks corresponding to the discrete nature of the system in the strong coupling regime. Those peaks practically disappear at $z=0.08$ where the curve of $\langle \Phi(u,z) \rangle$ tends to a flat region $[0,R_1]$ corresponding to the inner layer. This is still consistent with electrostatics since the inner layer is metallic and it corresponds to an equipotential surface. The average potential evaluated near the outer layer $[R_2,R_3]$ also exhibits peaks for $z=0.004$ and $z=0.04$. The number of peaks on the potential profile Fig.~\ref{StrongCouplingFig}-c \& d corresponds to the number of rings of the typical configurations near the ground state (one of them plotted in the inset image of Fig.~\ref{StrongCouplingFig}-d) and the stair-like profile of $q/Q$ at the outer layer. The behavior of the system at the strong coupling regime is therefore not described by the standard electrostatics due to the discrete effects.  As the $\beta$ parameter is increased, the circular arrangements of particles on the outer (typical of the ground state of the finite system) disappear due to the decoupling among particles. The progressing growth of the $\beta$ parameter also affects the thermodynamic average of the potential evaluated near the outer layer, as it is shown in Fig.~\ref{betaBehaviourOuterLayerFig}. In that figure, one can observe that $<\phi(z,u)>$ tends to be flat as it is evaluated at $z=z_1=R_1/100$ near the outer metallic layer. This is according to electrostatics since metallic layers are equipotential surfaces. We have to remark that $<\phi(z,u)>$ starts to be an equipotential function near both metallic layers if particles move enough to demonstrate a velocity distribution (consistent with the Maxwell-Boltzmann distribution) such as the one in Eq.~\ref{MaxwellBoltzmannDistribuitonEq} (the one of the ideal gas), but sufficiently electrically coupled to maintain the electrostatic features.   

\begin{figure}[H]
\centering 
\includegraphics[width=0.4\textwidth]{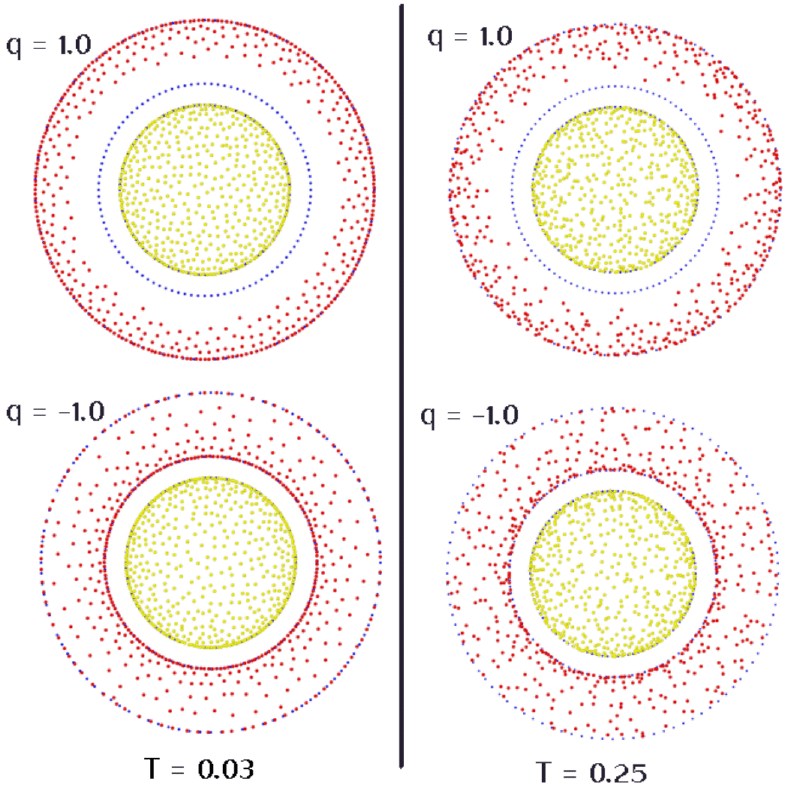}
\includegraphics[width=0.58\textwidth]{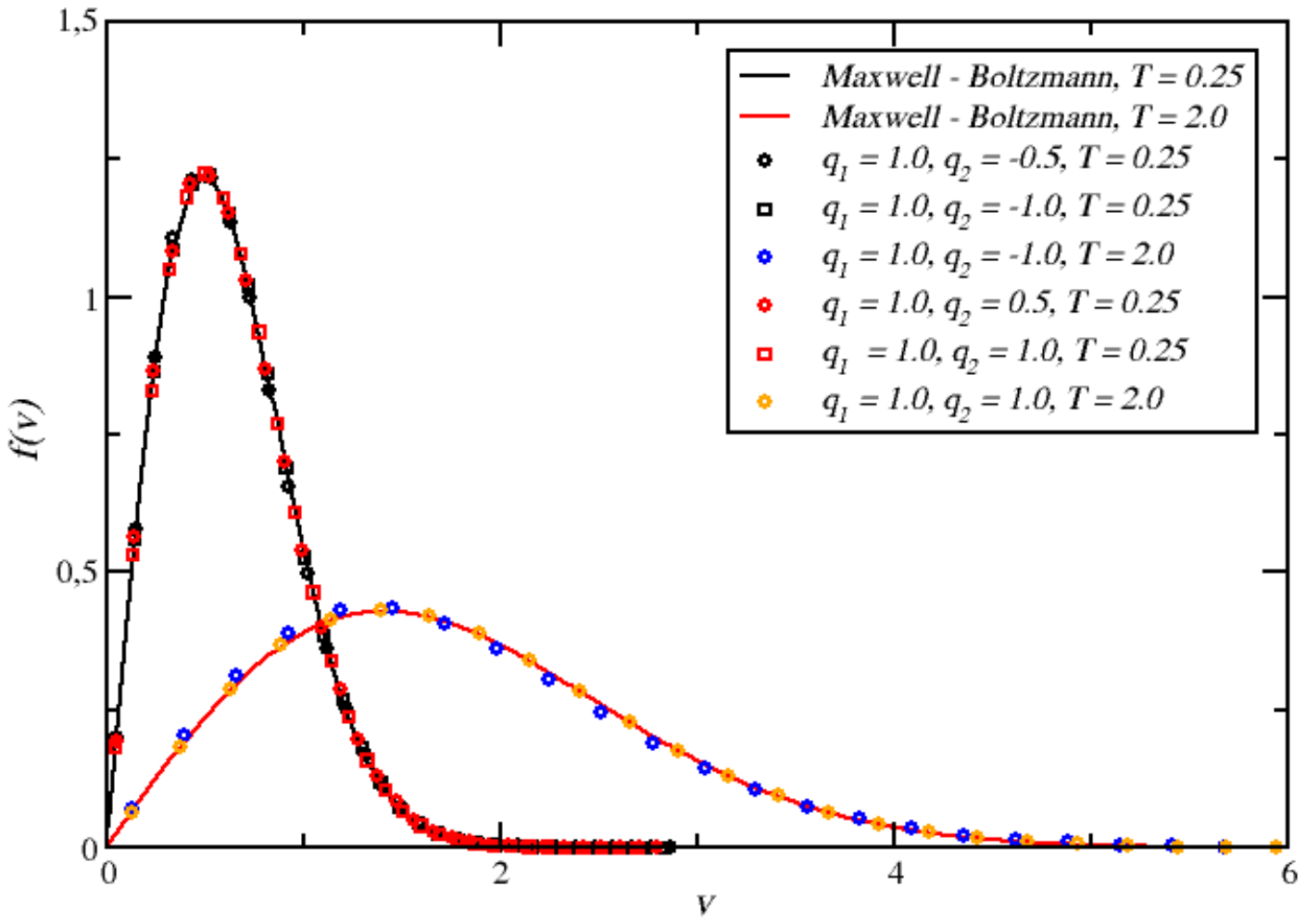}
    \caption[Total charge.]{MD velocity distribution for different couplings.   }
\label{MaxwellVelocityFig}
\end{figure}

\begin{figure}[H]
\centering 
\includegraphics[width=0.45\textwidth]{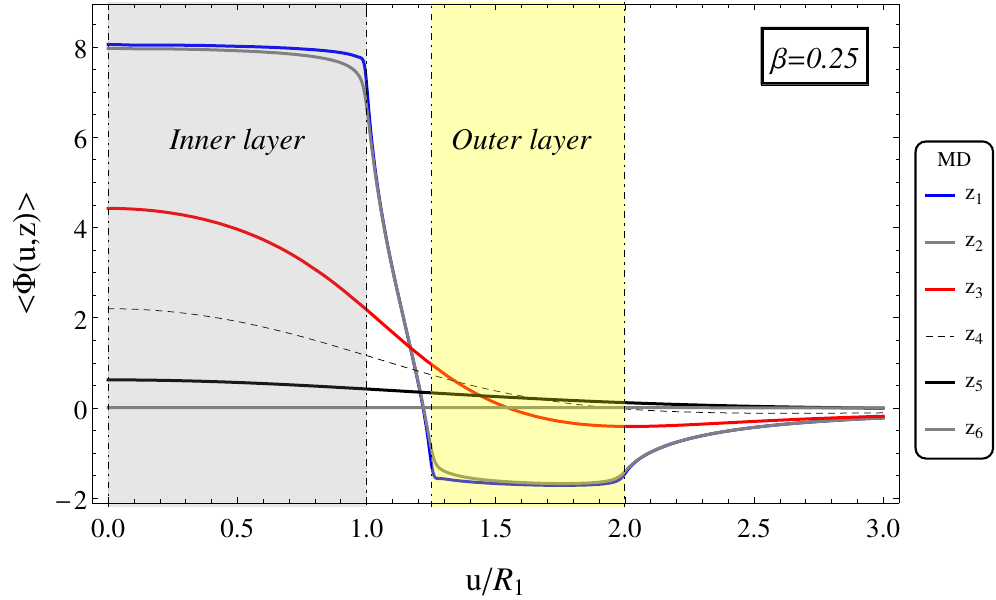}
\includegraphics[width=0.45\textwidth]{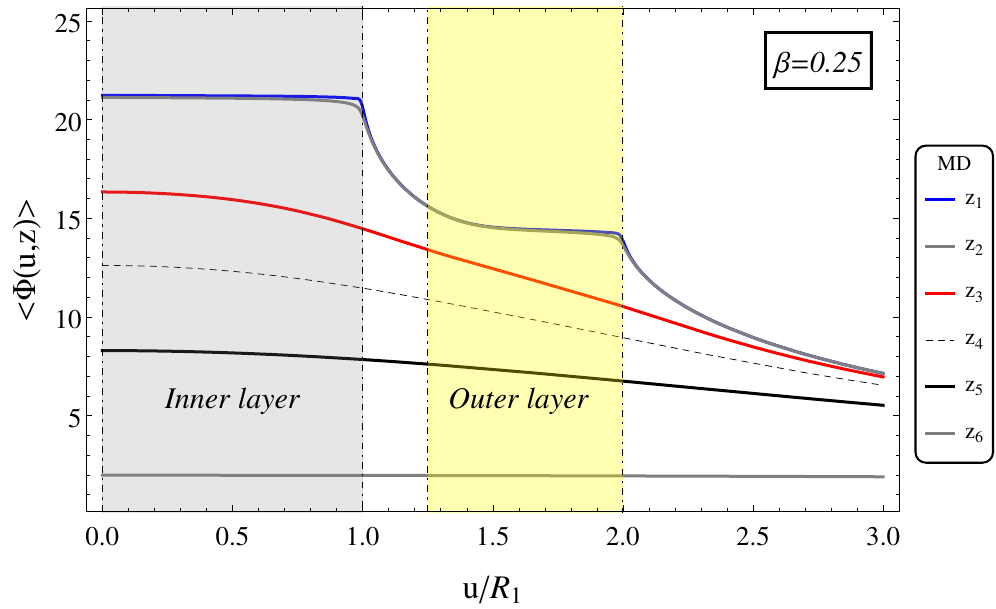}\\
\includegraphics[width=0.45\textwidth]{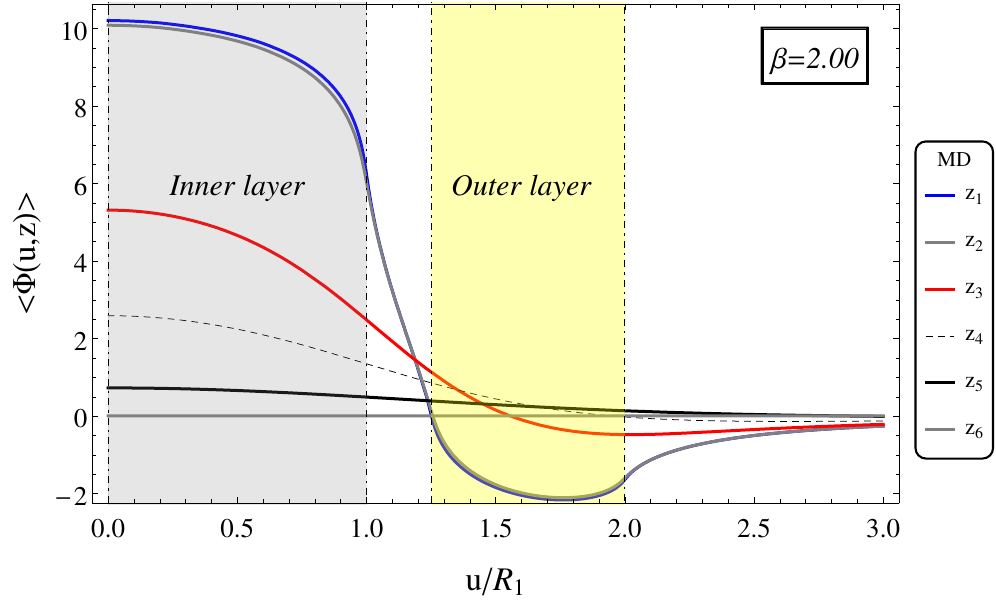}
\includegraphics[width=0.45\textwidth]{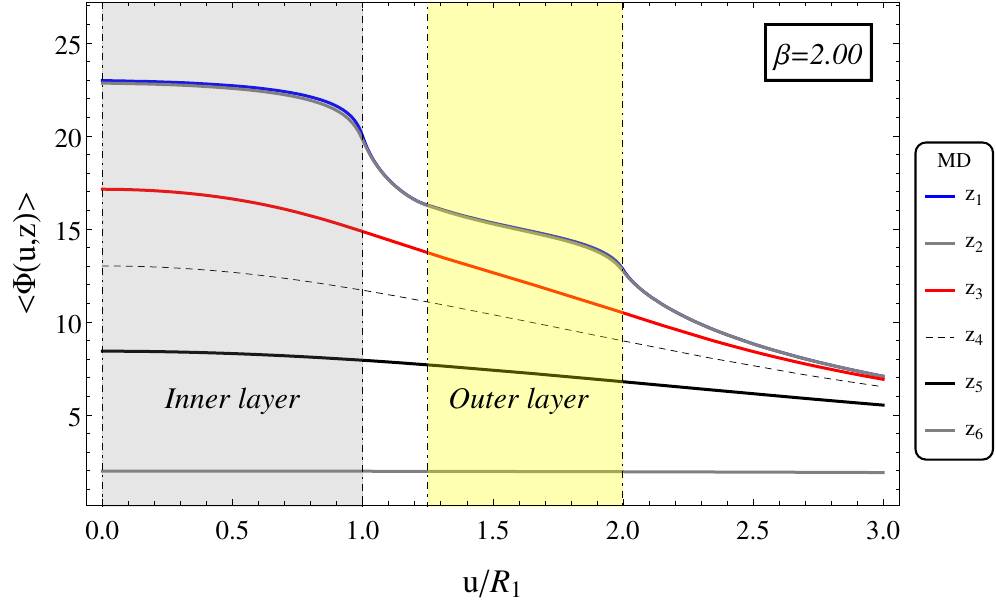}\\
\includegraphics[width=0.45\textwidth]{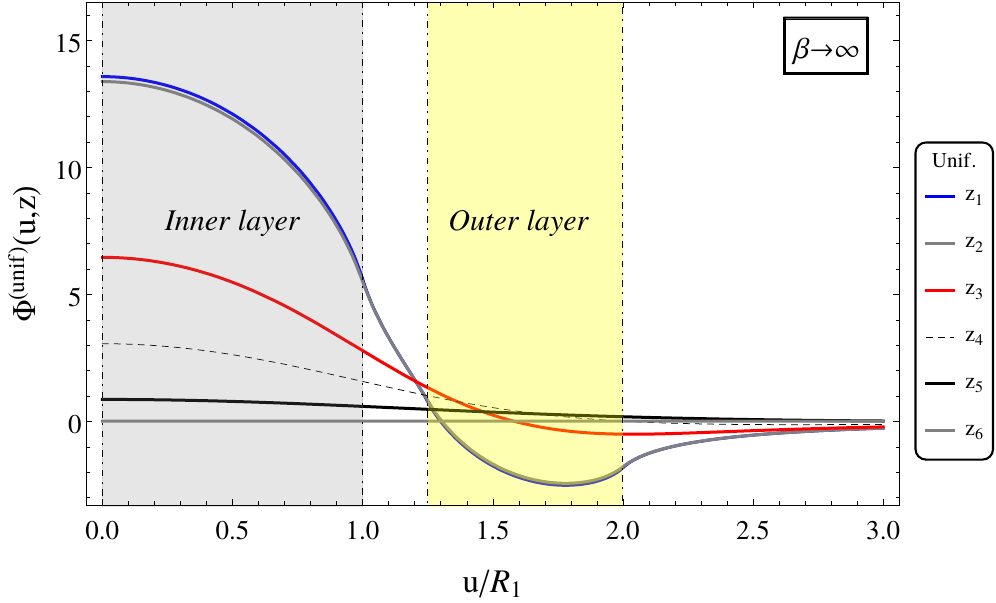}
\includegraphics[width=0.45\textwidth]{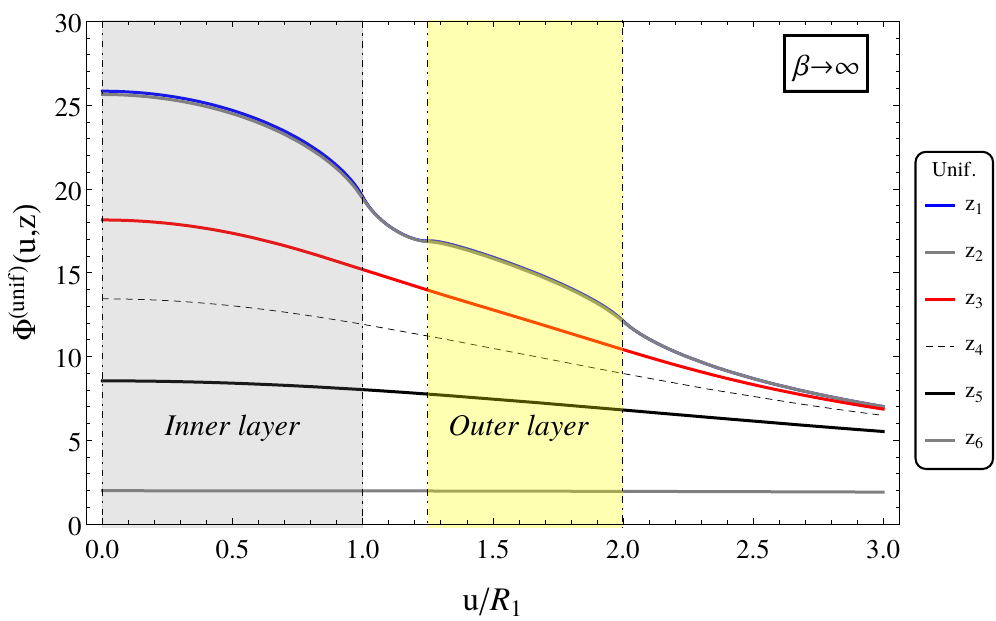}
    \caption[Total charge.]{MD potential profiles as a function of $\beta$. (left) Attractive layers $\xi=-1$ and (right) repulsive layers $\xi=1$.   }
\label{PotentialsBeta2Fig}
\end{figure}

In Fig.~\ref{MaxwellVelocityFig}, the MD velocity distribution is shown for $\beta = 0.25$ and $\beta = 2.0$, and charge ratios of $\xi=1$ (repulsive layers) and $\xi=-1$ (attractive layers). In both cases the MD speed distribution $f(v)$ computed with the Nosé-Hoover thermostat is according to the Maxwell speed distribution given by Eq.~(\ref{MaxwellBoltzmannDistribuitonEq}). However, only the value $\beta = 0.25$  ensures that moving particles are sufficiently coupled to generate collectively a flat enough potential $<\phi(z,u)>$ on the layers, as it is shown in Fig.~\ref{PotentialsBeta2Fig} for attractive and repulsive layers. On the other hand, if the parameter is set as $\beta = 2.0$, then the thermodynamic average of the potential does not generate equipotentials near the layers (see Fig.~\ref{PotentialsBeta2Fig}).

The $\beta \rightarrow \infty$ limit case corresponds to a plasma, where kinetic energy greatly exceeds the electric interaction and the surface charge density on the layers must be uniform. A close study of the $\beta \rightarrow \infty$ limit for the single circular layer starts from the surface charge density,
\[
<\sigma(\boldsymbol{r})> = <n(\boldsymbol{r})> q = \frac{1}{Z_{N,\beta}}\frac{q}{(N-1)!}\prod_{n=2}^N \int_{0}^{2\pi} d\phi_n\int_0^{R} u_n du_n  \prod_{1\leq i<j \leq N}\exp\left(-\frac{1}{4\pi\epsilon_o}\frac{q_i q_j/\beta}{\sqrt{u_i^2+u_j^2-2u_i u_j\cos(\phi_{ij}) }}\right),
\]
with $q$ the charge of the particles (assumed to be identical), $R$ the radius of the layer, $\phi_{ij}=\phi_i-\phi_j$, and  
\[
Z_{N,\beta} = \frac{1}{N!}\prod_{n=1}^N \int_{0}^{2\pi} d\phi_n\int_0^{R} u_n du_n  \prod_{1\leq i<j \leq N}\exp\left(-\frac{1}{4\pi\epsilon_o}\frac{q_i q_j/\beta}{\sqrt{u_i^2+u_j^2-2u_i u_j\cos(\phi_{ij}) }}\right)
\]
the canonical partition function. In the $\beta\rightarrow\infty$ limit, the partition function simplifies to a two-dimensional ideal gas and
\[
\lim_{\beta \to \infty} <\sigma(\boldsymbol{r})> = \frac{N q}{\pi R^2} 
\]
is distributed uniformly on the layer. The same occurs for the two-layer system. Then, the potential can be developed from
\begin{align*}
\lim_{\beta \to \infty}\langle\Phi(\boldsymbol{r})\rangle &= \frac{1}{4\pi\epsilon_o} \int_{\Omega_{in}\cup\Omega_{out}} \lim_{\beta \to \infty}\frac{\langle\sigma(\boldsymbol{r}')\rangle}{|\boldsymbol{r}-\boldsymbol{r}'|}d^2\boldsymbol{r} \\   
&=\frac{1}{4\pi\epsilon_o} \int_{u'\in [0, R_1]\cup[R_2, R_3]} \lim_{\beta \to \infty}\langle\sigma(u')\rangle u'du' \int_{0}^{2\pi}  \frac{d\phi'}{ \sqrt{u^2+u'^2 - 2uu'\cos(\phi-\phi') + z^2}} \\
&=\frac{1}{4\pi\epsilon_o} \int_{u'\in [0, R_1]\cup[R_2, R_3]} \lim_{\beta \to \infty}\langle\sigma(u')\rangle u'du' \frac{4}{\sqrt{(u-u')^2+z^2}}K\left(-\frac{4uu'}{(u-u')^2+z^2}\right).
\end{align*}
Since the surface charge density at each layer is uniform at the $\beta \to \infty $ limit, then
\begin{equation}
\Phi^{(unif)}(u,z) = \lim_{\beta \to \infty}\langle\Phi(\boldsymbol{r})\rangle = \frac{1}{\pi\epsilon_o} \sum_{k=1,2} \frac{N_k q_k}{A_k} \int_{U_k}  \frac{u' d u'}{\sqrt{(u-u')^2+z^2}}K\left(-\frac{4uu'}{(u-u')^2+z^2}\right),    
\label{PHIUniformEq}
\end{equation}
with $U_1=[0,R_1]$, $U_1=[R_2,R_3]$, $A_1 = \pi R_1^2$, $A_2 = \pi (R_3^2-R_2^2)$ the area of the layers. A plot of Eq~(\ref{PHIUniformEq}) is shown in Fig.~\ref{PotentialsBeta2Fig}. We can observe that MD simulations for $\beta=2.0$ (far from the thermodynamic limit) and Eq.~(\ref{PHIUniformEq}) for $\beta\to\infty$ do not generate equipotential surfaces since the surface charge distribution tends to be uniform, as occurs with the ideal gas.


\subsection{The gapless limit}
\label{gaplessLimitLbl}

The exact density charge in the gapless limit is given by 

\begin{equation}
\sigma^{Gapless}(u) = \lim_{R_3 \to \infty}\lim_{R_2 \to R_1} \sigma(u) = \frac{2\epsilon_o (V_{in}-V_{out})}{\pi} \left\{\frac{1}{R_1-u} E\left[\frac{4R_1 u}{(R_1+u)^2}\right] + \frac{1}{R_1+u} K\left[\frac{4R_1 u}{(R_1+u)^2}\right] \right\},
\label{sigmaGaplessEq}
\end{equation}

with $K(z)$ and $E(z)$ the complete elliptic integrals of the first and second kind respectively. Eq.~(\ref{sigmaGaplessEq}) can be obtained by using an electric vector potential approach described in \cite{salazar2020gaped, salazar2019AngularDependentSE}. This limit is achieved in MD when we set $N_1=N_2$ and $\xi = -1$ e.g. the same number of particles in both layers with opposite signs. 

\begin{figure}[H]
\centering 
\includegraphics[width=1.0\textwidth]{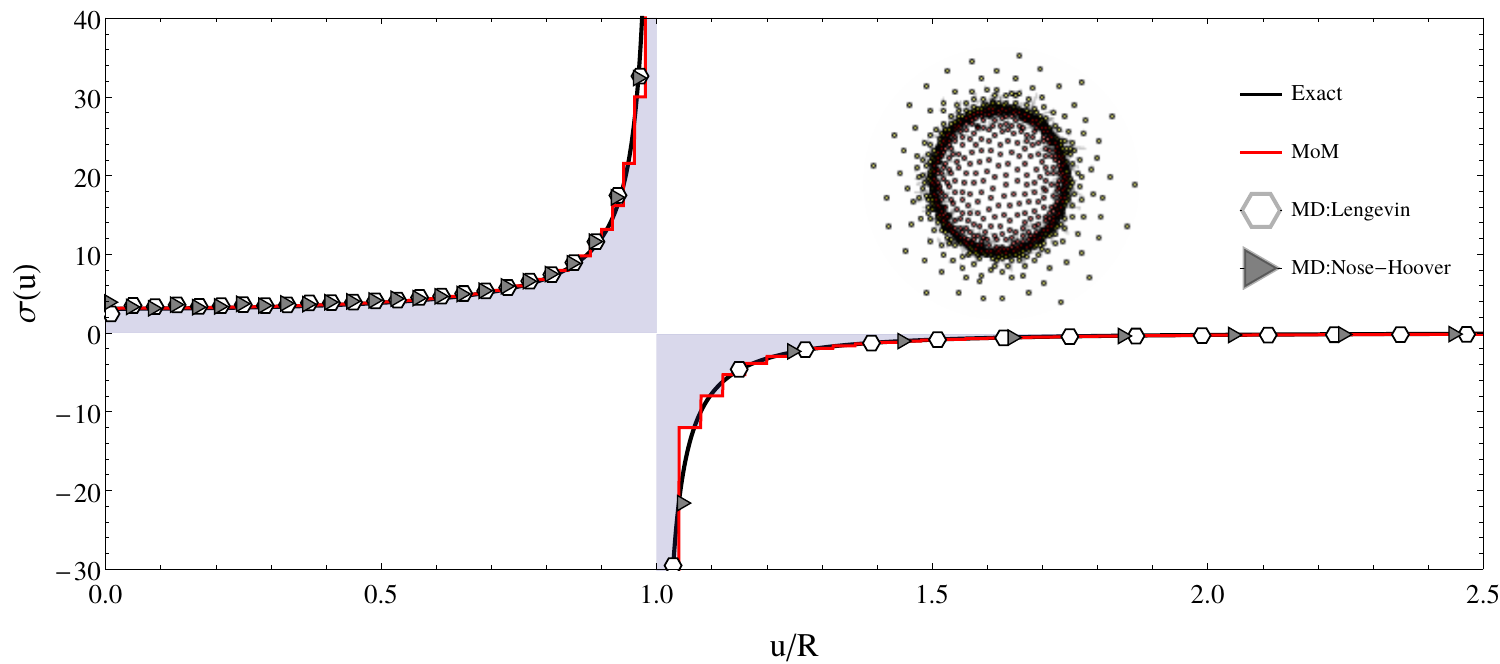}
    \caption[Total charge.]{Surface density charge in the gapless limit. The inner sheet is at  $V_{in} = 6.2043V$ and the outer one is grounded $V_{out}=-0.1028V$. The exact solution corresponds to Eq.~(\ref{sigmaGaplessEq}). MD simulations are performed with $N=1000$ particles at $\beta=0.25$. The inset figure corresponds to a equilibrated configuration of the particles in both layers via MD simulations.}
\label{sigmaGaplessFig}
\end{figure}

One can use Eq.~(\ref{sigmaGaplessEq}) as a benchmark for the current problem in the gapless limit. Fig.~\ref{sigmaGaplessFig} compares the density profiles due to MoM and the exact solution for $R_1 = 0.999 R_2$ and $R_3=3R_2$. Note that, formally, the Eq.~(\ref{sigmaGaplessEq}) works in the  $R_3\rightarrow\infty$ limit. Nonetheless, the MoM gives an accurate approximation for a finite value of $R_3$, since particles of different charges are strongly coupled near the gap as $R_1\rightarrow R_2$ and practically no particle reaches the boundary at $R_3$. Similarly, the MD results for the density profile calculated with the Langevin and Nose-Hoover thermostats are also in agreement with the exact solution, as shown in Fig.~\ref{sigmaGaplessFig}. Note that Eq.~(\ref{sigmaGaplessEq}) strictly works for $\xi=-1$, this is, opposite charges in each layer where the system is globally neutral. In this scenario, charges on the metallic layers are highly concentrated near the $R=(R_1+R_2)/2$ circle due to electric attraction (see the inset image in Fig.~\ref{sigmaGaplessFig}). Other scenarios concerning equal charge as $\xi=1$ cannot be modeled in the gapless limit since particles on the outer layer will escape to the infinite.

\subsection{The gapped case}

This subsection presents the results for the finite SE system with a gap. The parameters which define the geometry were set as follows: $R_1=4$, $R_2=5 R_1/4$ and $R_3=2R_1$. 
\begin{figure}[H]
\centering 
\includegraphics[width=0.85\textwidth]{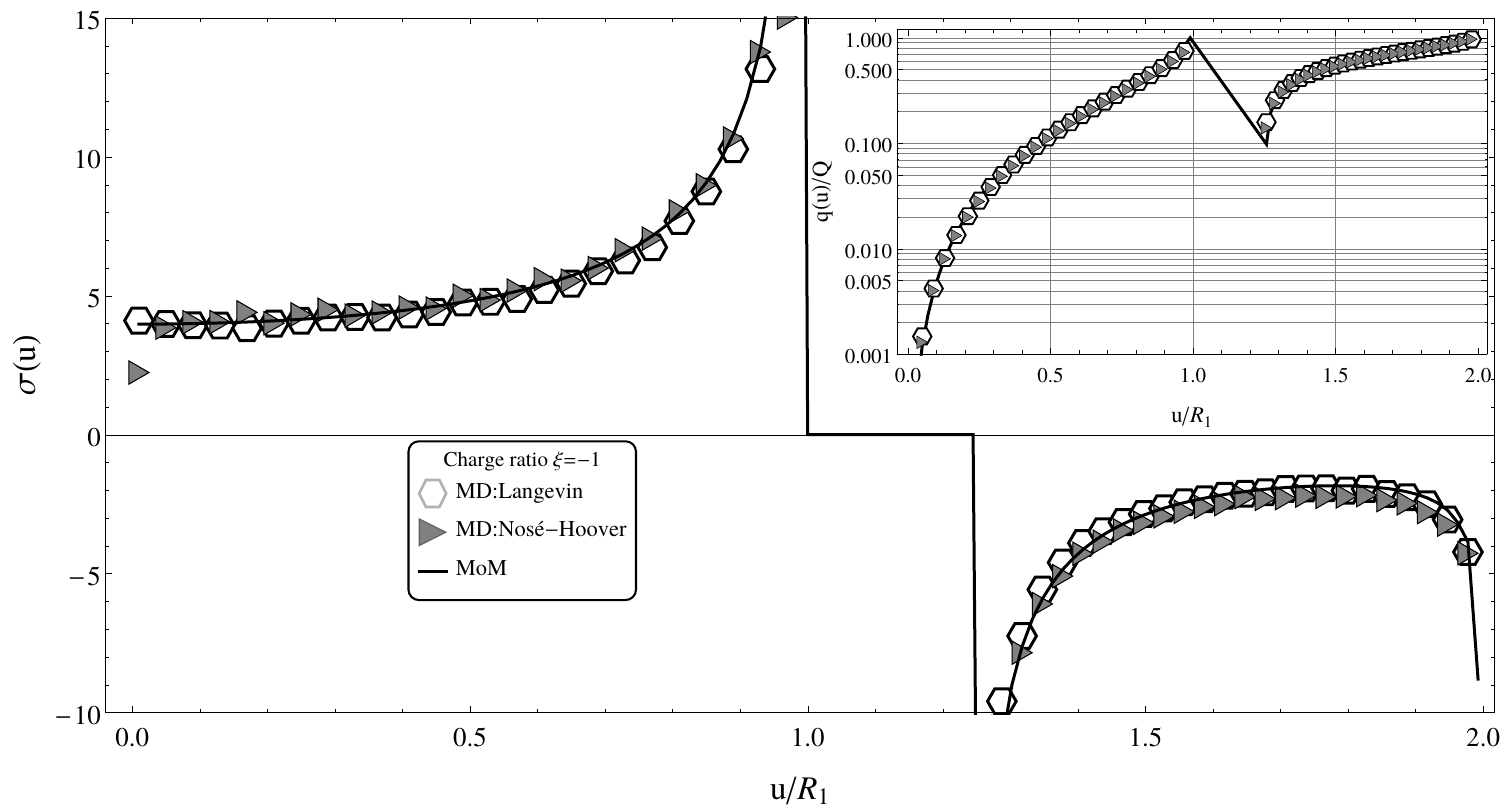}
    \caption[Total charge.]{Surface density charge for $\xi=-1$ and $\beta=0.25$. Symbols correspond to MD simulations and the solid line represents the MoM solution. }
\label{sigmaXiMinus1Fig}
\end{figure} 

The plot in Fig.~\ref{sigmaXiMinus1Fig} shows the density profile on both sheets for $\xi=-1$, this is, inner particles having positive charge $q_1=q$ and outer ones carrying opposite charge $q_1=-q$ thus the system being globally neutral. The solid line in Fig.~\ref{sigmaXiMinus1Fig} corresponds to the MoM approximation calculated according to Eq.~(\ref{MoMSigmaTwoSheetsEq}). MoM solution assumes that the density charge is continuous and it differs from the one in Eq.~(\ref{sigmaGaplessEq}) (valid in the gapless limit) since MoM predicts a charge concentration in the $R_3$ boundary. One observes that numerical solutions provided by MD simulations with Langevin and Nosé-Hoover thermostats are in agreement with the MoM solution. The parameter $\beta$ in the MD simulation is set as 0.3 for both thermostats. The inset semi log-plot in Fig.~\ref{sigmaXiMinus1Fig} is the integrated charge (see Eq.~(\ref{integratedChargeEq})), this is, the total charge stored in a circular region of radius $u$ in each layer scaled by the total charge $Q$ of the layer.  

\begin{figure}[H]
\centering 
\includegraphics[width=0.8\textwidth]{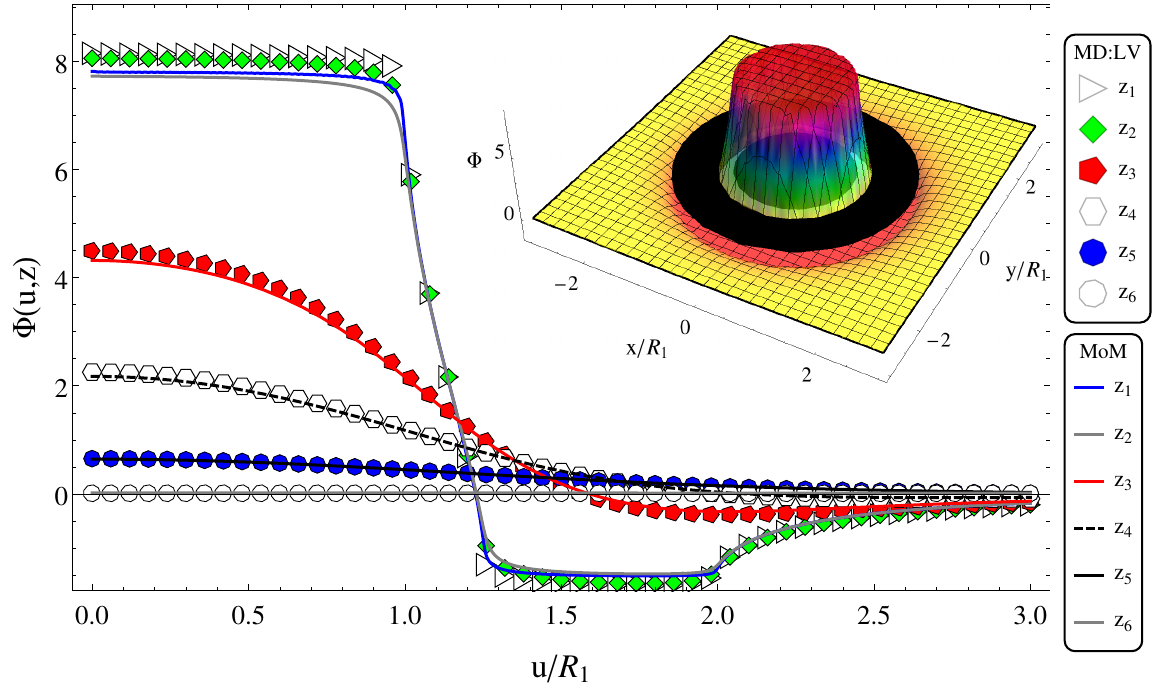}
    \caption[Total charge.]{Scalar electric potential for $\xi=-1$ and $\beta=0.25$. The inset plot is the potential in the $\mathbb{R}^3$-space close to the $xy$-plane at $z=z_1$. Metallic layers are represented as black surfaces. The symbols are the MD simulations with the Langevin thermostat and solid lines correspond to MoM solution. The set of heights chosen for this plot are $z_1=R_1/100$, $z_2=R_1/50$, $z_3=R_1/2$, $z_4=R_1$,$z_5=2R_1$, and $z_6=10R_1$.}
\label{potentialXiMinus1Fig}
\end{figure} 

Fig.~\ref{potentialXiMinus1Fig} shows the scalar electric potential profiles due to the charged layers at different heights $z_1<z_2<\ldots<z_6$. Methodologically we start by running a MD simulation with a fixed geometry, number of particles, and $\beta$ parameter. Later the surface charge density is calculated by thermodynamic averages of the number density of each layer. We build the scalar potential profile very close to the $xy$-plane by using Eq.~(\ref{thermodynamicAvergaPhiEq}), which enables us to compute the inner $V_{in}$ and outer $V_{in}$ voltages. The voltages computed via MD simulations are used as input parameters in Eq.~(\ref{MoMSigmaTwoSheetsEq}) to compute surface charge density with the MoM. That procedure not only ensures that both numerical methods generate profiles on the same scale but also the same profile distributions since a random choice of voltages in the MoM may develop profiles corresponding to several ratios of charges.

\begin{figure}[H]
\centering 
\includegraphics[width=0.8\textwidth]{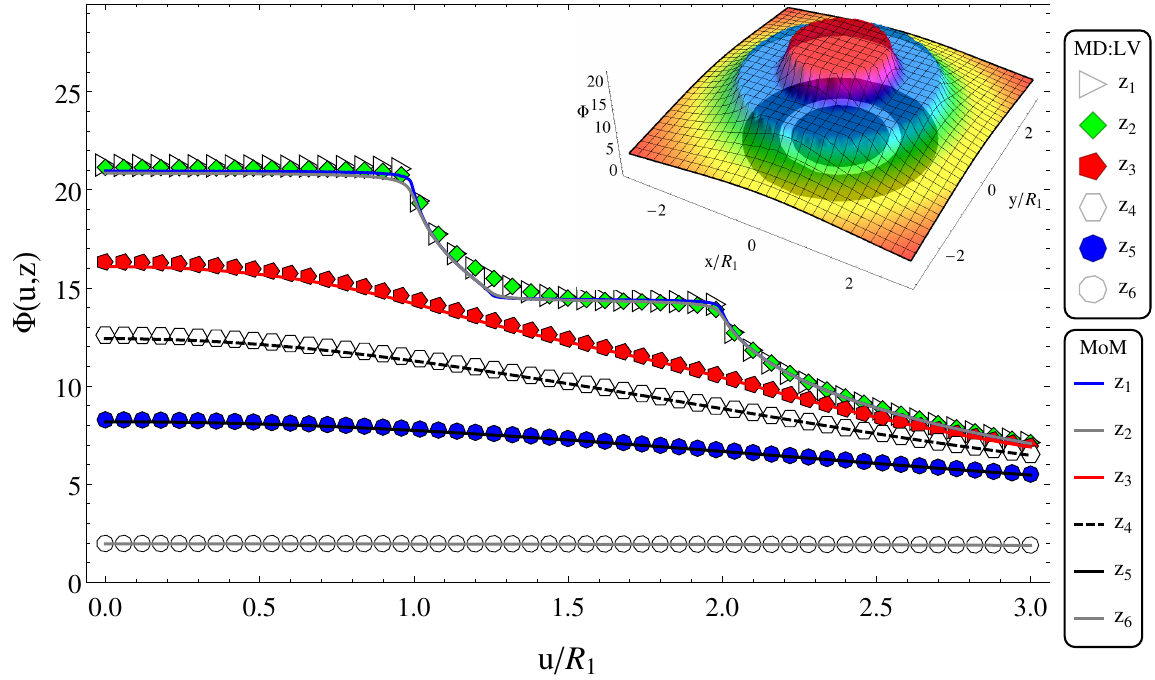}
    \caption[Total charge.]{Scalar electric potential for $\xi=1$ and $\beta=0.25$. The inset plot is the potential in the $\mathbb{R}^3$-space close to the $xy$-plane at $z=z_1$. Metallic layers are represented as black surfaces. The symbols are the MD simulations with the Langevin thermostat and solid lines correspond to MoM solution.}
\label{potentialXiPlus1Fig}
\end{figure}

Fig.~\ref{potentialXiPlus1Fig} shows the scalar electric potential for the $\xi=1$ case, this is, the same charge on both layers. The MD simulations are performed with a thousand of identically charged particles, $N_1=500$ charges in the inner layer and $N_2=500$ in the outer one. In general, the results of the DM potential and the MoM are also consistent for the case of repulsive layers. However, unlike the attractive case, there is a slight discrepancy in the potential over the outer plate near the edge $R_3$ as can be observed in Fig. 10, where $z = z_1$. This occurs because the particles on the inner plate, having the same charge as the particles on the outer plate, repel the latter in such a way that these charges less frequently visit the boundary $R = R_3$. This results in the statistics near this boundary not being as good and the potential appearing less constant in this region of the conductor.

\begin{figure}[H]
\centering 
\includegraphics[width=0.45\textwidth]{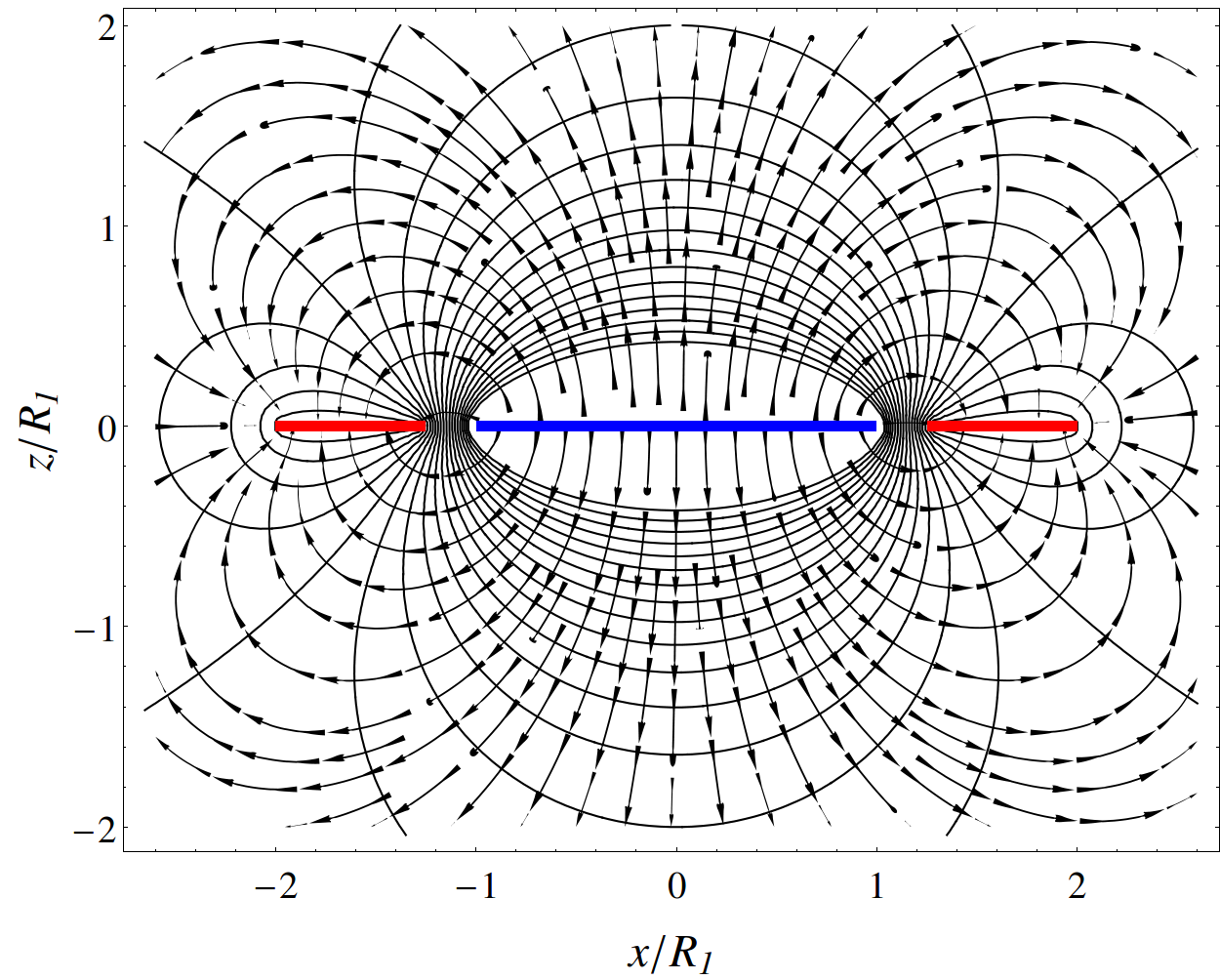}
\includegraphics[width=0.45\textwidth]{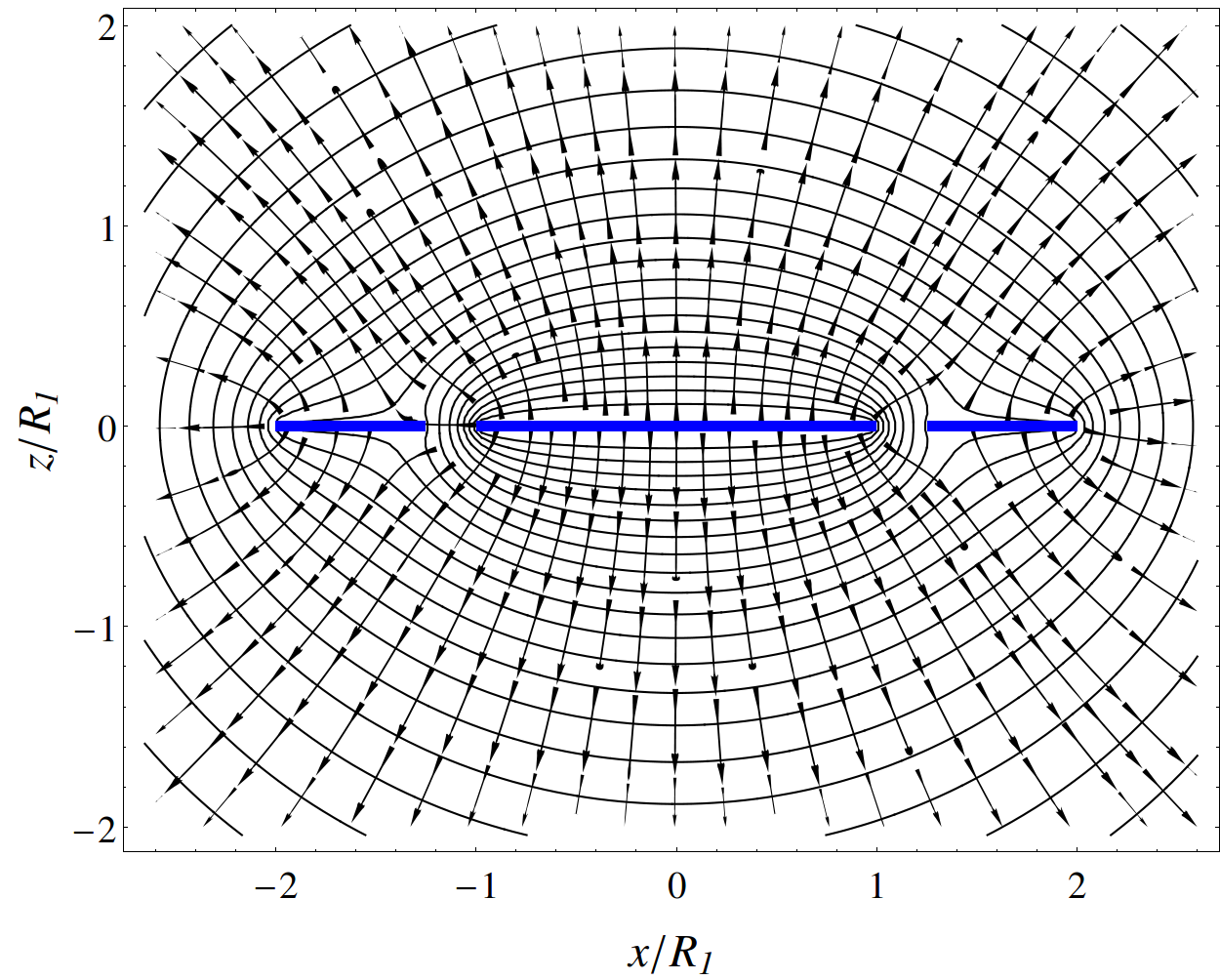}
    \caption[Total charge.]{Electric field from the thermodynamic average $<\sigma>$ at $\beta=0.25$ computed via MD simulations. The field is evaluated on the plane $y=0$. (left) Layers having opposite charges and (right) the same charge. Metallic layers are represented as solid lines on the $x$-axis. The blue and red color corresponds to the positive and negative charge respectively.  }
\label{fieldsFig}
\end{figure}

The electric field due to configurations for $\xi=-1$ (opposite charge) and $\xi=1$ (equal charge) are shown in Fig.~\ref{fieldsFig}. The electric field is obtained from computing the numerical gradient of Eq.~(\ref{thermodynamicAvergaPhiEq}). That is,
\[
E_u(u,z) = - \frac{\partial}{\partial u} \langle \Phi(u,z) \rangle = - \lim_{\Delta u \to 0} \frac{ \langle \Phi(u+\Delta u,z) \rangle -  \langle \Phi(u,z) \rangle}{\Delta u}
\]
and
\[
E_z(u,z) = -\frac{\partial}{\partial z} \langle \Phi(u,z) \rangle  = - \lim_{\Delta z \to 0} \frac{ \langle \Phi(u,z+\Delta z) \rangle -  \langle \Phi(u,z) \rangle}{\Delta z}
\]
by the forward finite differences of the scalar potential thermodynamic average. MD thermodynamic averages of the electric field agree with the electrostatic prediction of the streamlines of $\boldsymbol{E}$. Thus we can observe agreement between the discrete MD and continuous MoM approaches at $\beta=0.25$ since thermodynamic averages of the electric potential near the layer tend to a constant value (see Figs.~\ref{potentialXiMinus1Fig} and Figs.~\ref{potentialXiPlus1Fig}), and consequently, the electric field streamlines become orthogonal to the layers (see Fig.~\ref{fieldsFig}).    

\section{Conclusions}
In this manuscript, we conducted molecular dynamics simulations of a classical plasma in two dimensions, incorporating Coulomb interactions within a circular geometry devoid of background density. The system's electrostatic analog is a circular surface electrode in the context of finite metallic layers held at fixed potentials. Our methodology emphasized the application of molecular dynamics simulations and the Method of Moments to solve the system, alongside developing a protocol for their comparative analysis. This rigorous approach revealed the MD method's efficacy in correctly describing the system under specific coupling regimes, highlighting the nuanced behavior of the two-dimensional plasma as opposed to the simpler SE system.


The complexity of the plasma system extends beyond the scope of continuum media electrostatics, especially when considering scenarios far from the thermodynamic limit. MD simulations near the ground state showed finite size effects leading to circular arrangements of particles, which significantly influence the thermodynamic average of potential near the metallic layers. Notably, a comparison of the integrated charge on the layers between MD and MoM at the strong coupling regime ($\beta \rightarrow 0$) showed a staircase-like behavior, absent in electrostatics, for which the integrated charge profile is smooth. Moreover, at the limit $\beta \rightarrow \infty$, the system generates uniform surface charge distributions, resulting in potential profiles on the layers that defy the equipotential surfaces inherent in electrostatics of metallic layers. 

Our numerical analysis determined that achieving the equipotential characteristic of metallic layers from the Coulomb plasma requires increasing the value of $\beta$ from the ground state. This discovery facilitated the identification of a $\beta$ range where the thermodynamic average of surface charge density via MD closely resembles that predicted by the MoM for the SE system. Within this range, particularly at $\beta=0.25$, finite-size effects of the Coulomb plasma, even with a thousand particles, become negligible in replicating the SE's electrostatics. This condition was further corroborated in the gapless limit, where MD simulations, employing Nosé-Hoover and Langevin thermostats, the MoM, and the analytical benchmark provided by Eq.~(\ref{sigmaGaplessEq}), demonstrated remarkable congruence.

While the gaped case lacks an exact solution, implementing the MoM allowed for the approximation of surface charge density on the metallic layers. The MoM's outcomes, particularly at $\beta=0.25$ and considering attractive layers ($\xi=-1$), aligned well with those derived from MD simulations. This comprehensive study not only underscores the critical role of methodological adherence in analyzing complex systems but also illuminates the intricate behavior of two-dimensional plasmas, thereby contributing valuable insights into the field of electrostatics and plasma physics.

\section*{Acknowledgments}
This work was partially supported by Vicerrector\'ia de investigaci\'on, Universidad ECCI through project IN-08-28. Camilo Bayona expresses gratitude to the Centro de Ingenier\'ia Avanzada, Investigaci\'on y Desarrollo - CIAID for the technical support provided for the present study. Cristian Cobos thanks the financial support of Centro de investigaci\'on e innovaci\'on en tecnolog\'ia y ciencia CEINTECCI project CEINTECCI-CPJI-2023-ID-14. Diego Jaramillo thanks Universidad Antonio Nari\~no for financial support through project No 2023202.

\bibliographystyle{ieeetr} 
\bibliography{bibliography.bib}

\newpage
\begin{appendices}
\section{The $1/r^{\eta}$ interaction}
\label{generalizationAppendixLbl}
Let us consider the electrostatic problem with the following particle-particle interaction potential $\Phi_{\eta}(r)=q\nu_{\eta}(r)$, with
\[
\nu_{\eta}(\boldsymbol{r},\boldsymbol{r'}) = \nu_{\eta}(|\boldsymbol{r}-\boldsymbol{r'}|) = \frac{L^{\eta-1}}{4\pi\epsilon_o}\frac{1}{|\boldsymbol{r}-\boldsymbol{r}'|^{\eta}}  
\]
a scalar function, $L$ a characteristic parameter with units of length, and $\eta$ a real positive number. The surface charge density is found from the following integral expression
\[
\int_{\Omega} \sigma(\boldsymbol{r}')\nu(\boldsymbol{r},\boldsymbol{r'}) d^2\boldsymbol{r} = \Phi(\boldsymbol{r}\in\Omega) = V_o,
\]
which can be written as 
\[
\sum_{n=0}^N \sigma_n \int_{0}^R \mathcal{I}_{\eta}(u_m^{(c)},u') f_n(u') u'du' = \Phi\left((u_m^{(c)},0,0)\in\Omega\right) = V_o.
\]
by introducing
\[
\mathcal{I}_{\eta}(u,u') = \frac{L^{\eta-1}}{4\pi\epsilon_o} \int_0^{2\pi} \nu_\eta(\sqrt{u^2+(u')^2 - 2uu'\cos(\beta)}) d \beta.
\]

\begin{figure}[H]
\centering 
\includegraphics[width=0.55\textwidth]{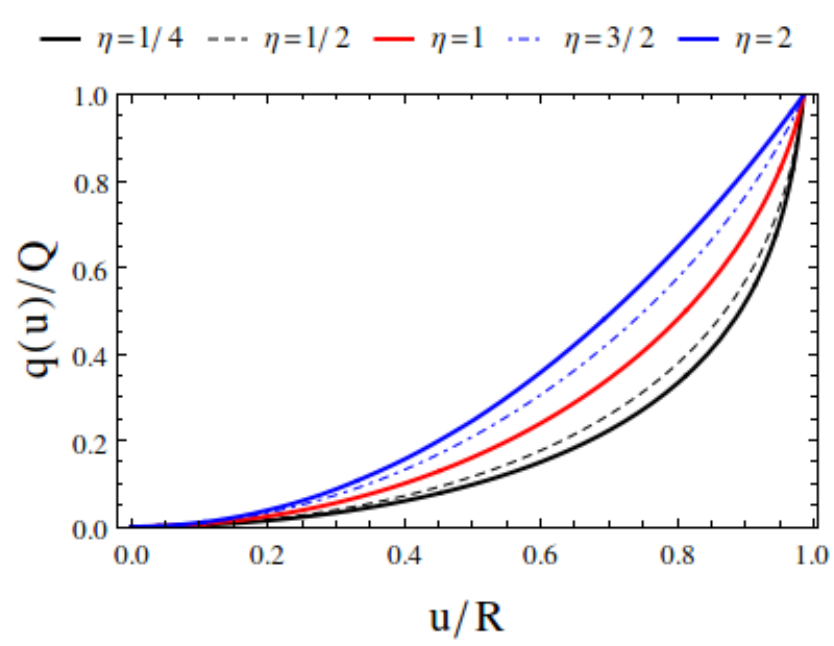}
    \caption[Total charge.]{Total charge.}
\label{totalChargeFig}
\end{figure} 

The angular integral in the previous expression can be written in terms of special functions, by following an analogous procedure than the one in Eq.~(\ref{auxAngularEq}). The angular integral is
\[
\int_0^{2\pi} \frac{d \beta}{(\sqrt{u^2+(u')^2 - 2uu'\cos(\beta)})^\eta}  = \frac{2\pi}{|u-u'|^\eta} {}_2 F_1\left( \frac{1}{2},\frac{\eta}{2}, 1; -\frac{4uu'}{|u-u'|^2} \right)
\]
where 
\[
{}_2F_1\left( a, b, c ; \xi \right) = \frac{\Gamma(c)}{\Gamma(b)\Gamma(c-b)} \int_{0}^1 \frac{t^{b-1}(1-t)^{c-b-1}}{(1-t\xi)^a}dt
\]
is Gauss hypergeometric function, and $\Gamma(z)$ is the gamma function. Once more, the problem can be reduced to a linear set of algebraic equations
\[
\sum_{n=0}^N M^{(\eta)}_{mn}\sigma_n = \Phi_m, 
\]
with
\[
M^{(\eta)}_{mn} =\int_{U_n} \mathcal{I}_{\eta}(u_m^{(c)},u') f_n(u') u'du' = \frac{L^{\eta-1}}{4\pi\epsilon_o} \int_{U_n} \left[\frac{2\pi}{|u_m^{(c)}-u'|^\eta} {}_2 F_1\left( \frac{1}{2},\frac{\eta}{2}, 1; -\frac{4u_m^{(c)}u'}{|u_m^{(c)}-u'|^2} \right)\right] f_n(u') u'du',
\]
being $U_n \in [0,R]$ the interval where the $n$-th basis function is different from zero. The surface charge density is shown in Fig.~\ref{sigmaFig} for several $\eta$ values.

Another quantity of interest is the charge inside a circular region, that can be computed from
\[
q(u) = 2\pi \int_0^u \sigma(u') du' = 2\pi \sum_{n=0}^N \sigma_n \int_{0}^u f_n(u')u'du',
\]
where the total charge is $Q=q(R)$.

\section{The Method of Moments for two circular layers}
\label{appMoMTwoLayersSec}
Let us generalize the MoM approach for two concentric layers on the plane. By studying the interplay between charges, we aim to elucidate the intricate correlation between the surface charge density and the underlying Coulomb interactions. 

\begin{figure}[H]
\centering 
\includegraphics[width=0.5\textwidth]{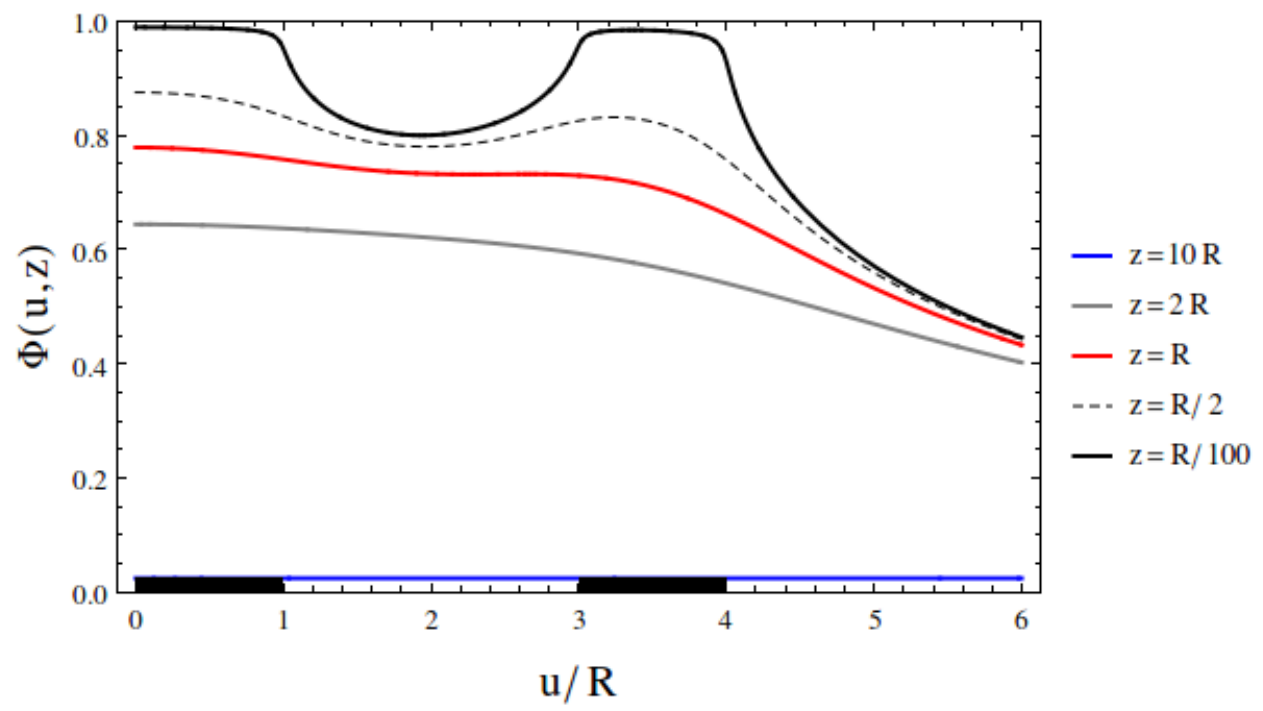}
    \caption[Total charge.]{Illustrative potential profiles at different heights. Both layers are at the same voltage $V_{in}=V_{out}=1V$ and their cross sections are represented with black rectangles. }
\label{potentialCurvesEqualIIFig}
\end{figure}

We proceed by writing the electrostatic integral formula for the scalar potential 
\[
\frac{1}{4\pi\epsilon_o} \int_{\Omega_{in}\cup\Omega_{out}} \frac{\sigma(\boldsymbol{r}')}{|\boldsymbol{r}-\boldsymbol{r}'|}d^2\boldsymbol{r} = \Phi(\boldsymbol{r}\in\Omega_{in}\cup\Omega_{out}) = \begin{cases}
    V_{in}   &\boldsymbol{r}\in\Omega_{in} \\
    V_{out}   &\boldsymbol{r}\in\Omega_{out}
\end{cases}
\]
with $V_{in}$ and $V_{out}$ the voltage of the inner and outer layers. One can use the angular symmetry of the system to write 
 \[
 \int_{u'\in [0, R_1]\cup[R_2, R_3]} \sigma(u') I(u,u') u'du' = \Phi(u,0,0) = \begin{cases}
    V_{in}   &u\in[0,R_1] \\
    V_{out}   &u\in[R_2,R_3]
\end{cases}
 \]
 with
\[
I(u,u')=\frac{1}{4\pi\epsilon_o}\int_{0}^{2\pi}  \frac{d\beta}{ \sqrt{u^2+u'^2 - 2uu'\cos(\beta)}} = \frac{1}{4\pi\epsilon_o}\frac{4}{|u-u'|}K\left(-\frac{4uu'}{|u-u'|^2}\right).
\]
The potential profiles at different heights of the system are shown in Fig.~\ref{potentialCurvesEqualIIFig}. We are interested in the surface density charge $\sigma$ on both layers. 

We define two basis: one having $\mathcal{N}$-basis functions on the inner layer $\left\{f_n^{(in)}(u)\right\}_{n=1,\ldots,\mathcal{N}}$ and another having $\mathcal{N}$-basis functions on the outer layer $\left\{f_n^{(out)}(u)\right\}_{n=\mathcal{N}+1,\ldots,2\mathcal{N}}$. Thus, the surface charge density can be computed from
\[
\sigma(u) = \sum_{n=1}^\mathcal{N}\sigma_n f_n^{(in)}(u) + \sum_{n=\mathcal{N}+1}^{2\mathcal{N}}\sigma_n f_n^{(out)}(u) = \sum_{n=1}^{2\mathcal{N}}\sigma_n f_n(u),
\]
with $f_{n}(u) = f_n^{(in)}(u) \hspace{0.25cm}\mbox{\textbf{if}}\hspace{0.25cm}1\leq n\leq \mathcal{N}, \hspace{0.25cm}\mbox{\textbf{otherwise}}\hspace{0.25cm}f_n^{(out)}(u)$. For simplicity, we shall take piece-wise functions as bases
\[
f_n^{in}(u)=\begin{cases}
    1 & u\in[u_{n-1},u_n] := U_{n}^{(in)} \\
    0 & u \notin U_{n}^{(in)}
\end{cases} \hspace{0.5cm}\mbox{and}\hspace{0.5cm}f_n^{out}(u)=\begin{cases}
    1 & u\in[u_{n},u_{n+1}] := U_{n}^{(out)} \\
    0 & u \notin U_{n}^{(out)}
\end{cases}
\]
where node points are 
\[
u_n=\begin{cases}
    u_n^{(in)}=n \frac{R_1}{\mathcal{N}} & n=0,\ldots,\mathcal{N}, \\
    u_n^{(out)}=R_2 + [n-(N+1)] \frac{R_3-R_2}{\mathcal{N}} & n=\mathcal{N}+1,\ldots,2\mathcal{N}+1  .
\end{cases}
\]

\begin{figure}[H]
\centering 
\includegraphics[width=0.22\textwidth]{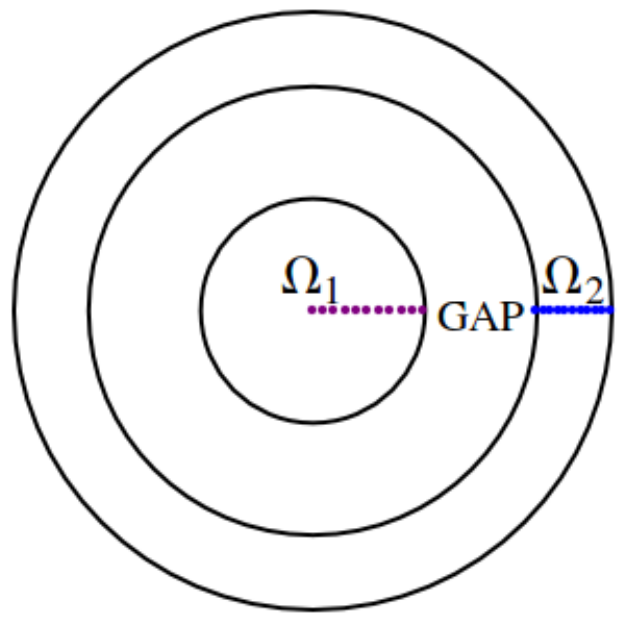}
\includegraphics[width=0.75\textwidth]{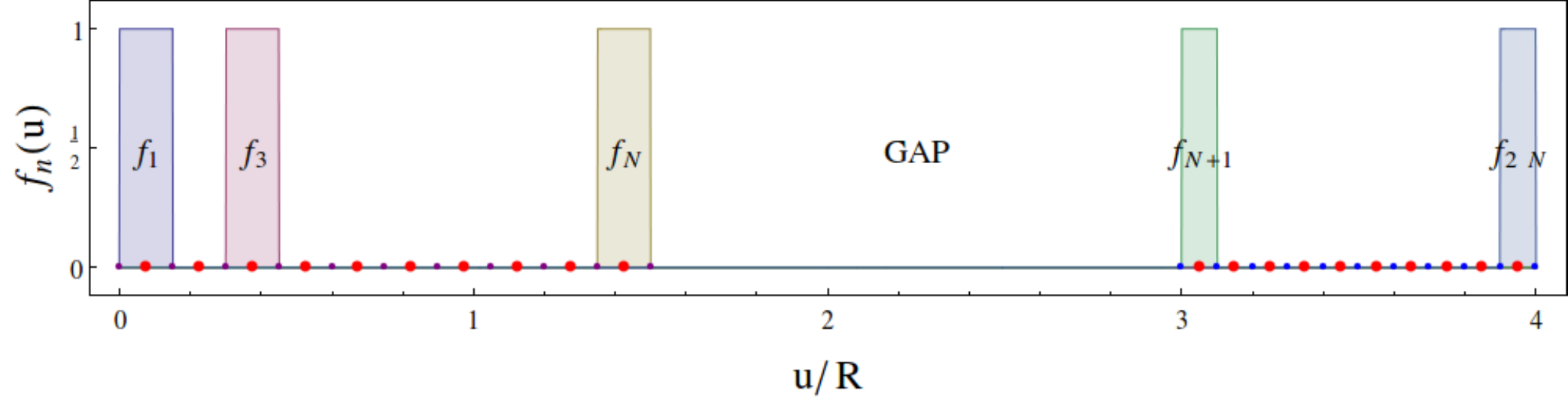}
    \caption[The basis.]{Illustrative discrete bases. (left) Layers and nodes $\{u_{n}\}_{n=1,\ldots,2\mathcal{N}+1}$ in purple color. (right) Basis functions. Red points are located at the element's center $\{u_{m}^{(c)}\}_{m=1,\ldots,2\mathcal{N}}$.   }
\label{basisFig}
\end{figure}

The basis functions are represented schematically in Fig.~\ref{basisFig}. Note that even when the bases are split into two groups of functions, simultaneously solving the problem for both layers $\Omega_1$ and $\Omega_2$ is required. 

\begin{figure}[H]
\centering 
\includegraphics[width=0.3\textwidth]{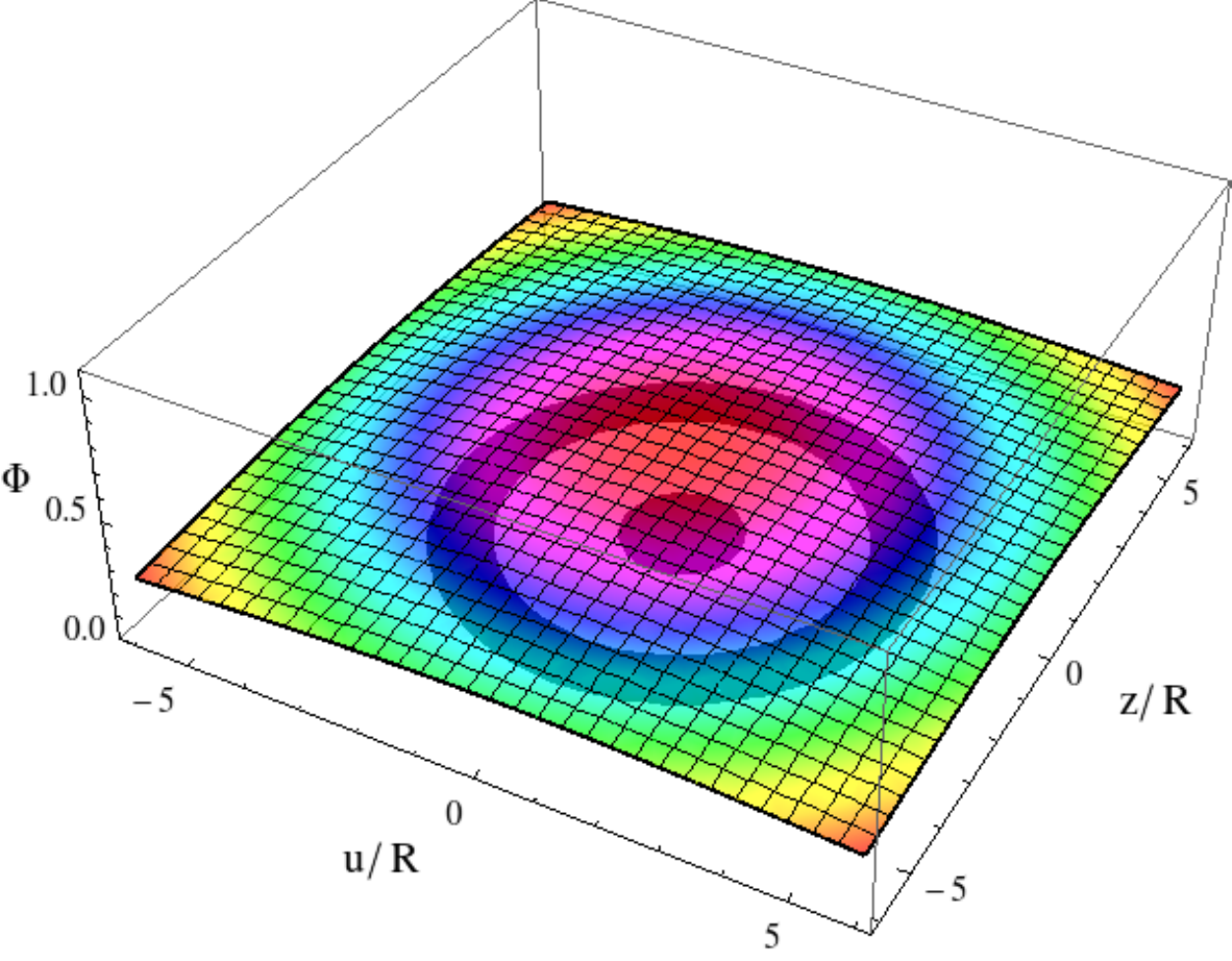}
\includegraphics[width=0.3\textwidth]{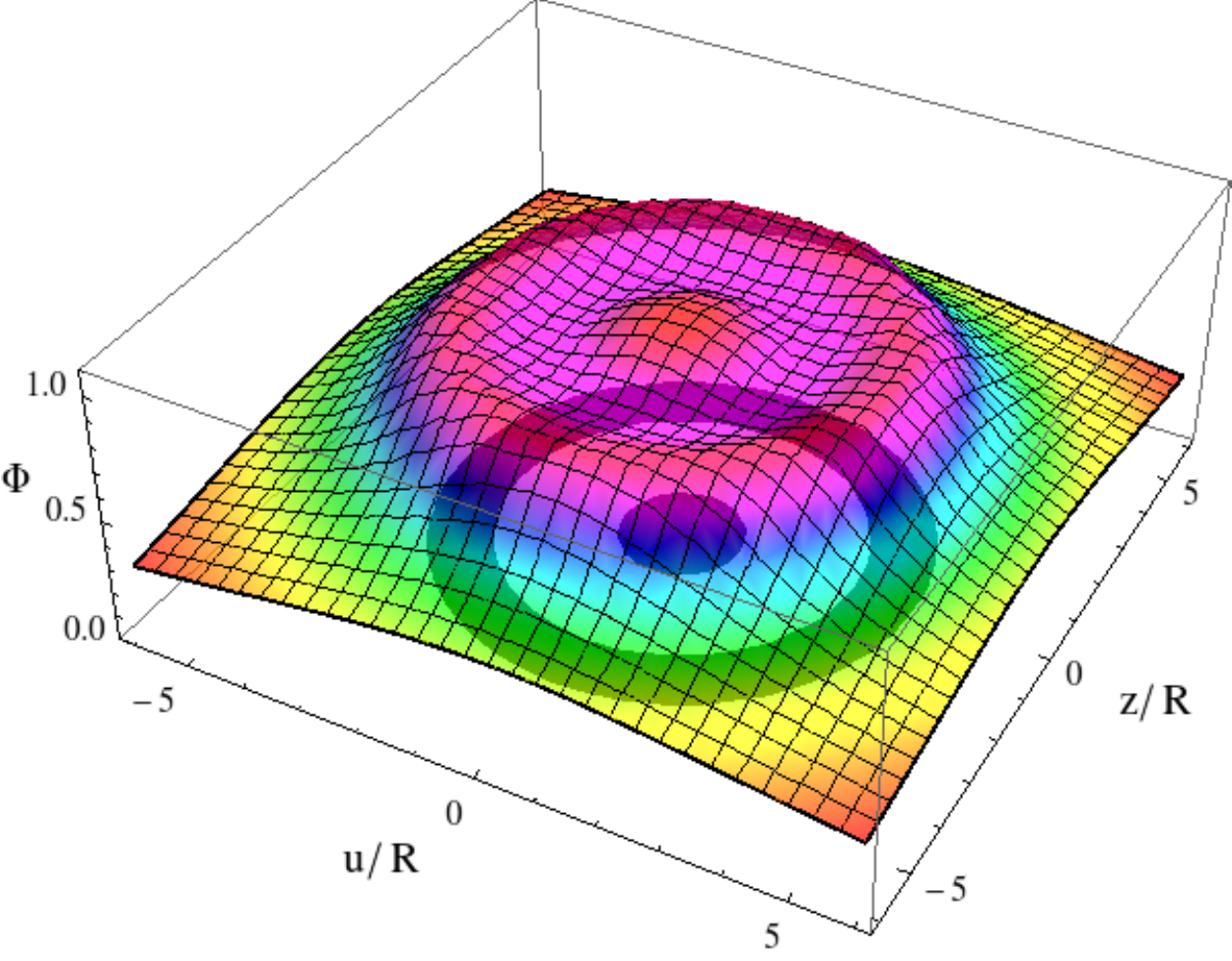}
\includegraphics[width=0.3\textwidth]{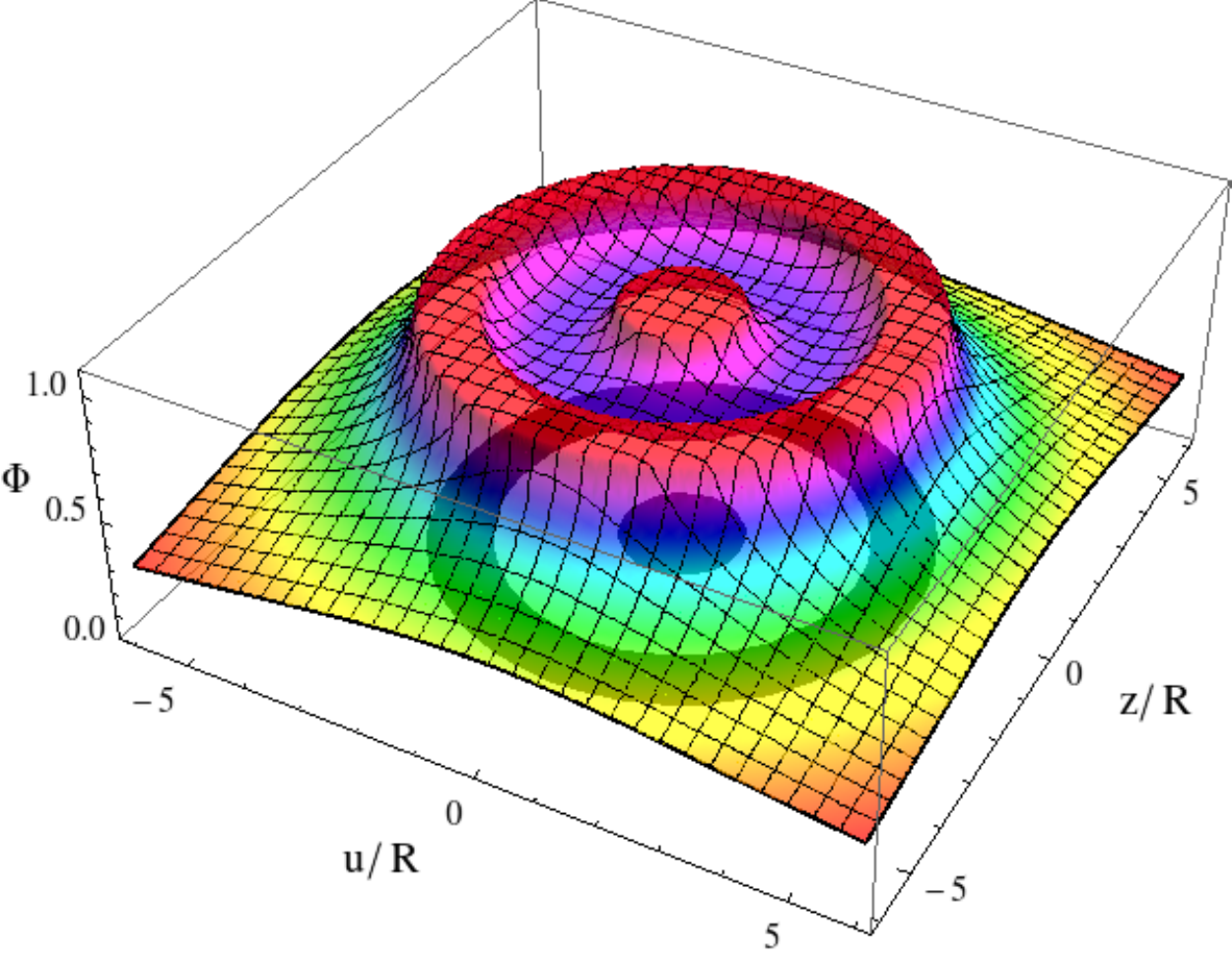}
    \caption[Total charge.]{Illustrative electric potential on the $z=$ constant plane. Potential $\Phi$ evaluated at $z=5R$, $z=0.3R$ and $z=0.025R$ (left to right). Both layers are set to the same voltage $V_{in}=V_{out}=1V$. }
\label{potentialSurfEqualFig}
\end{figure}

Therefore,
\[
 \sum_{n=1}^{2\mathcal{N}} \sigma_n \int_{u'\in [0, R_1]\cup[R_2, R_3]} f_{n}(u')  I(u,u') u'du' = \Phi(u,0,0) = \begin{cases}
    V_{in}   &u\in[0,R_1] \\
    V_{out}   &u\in[R_2,R_3]
\end{cases}
 \]
Since the radial distance $u$ is not constrained, we may evaluate the previous equation at the centers $\left\{u_m^{(c)}\right\}_{m=1,\ldots,2\mathcal{N}}$ of the elements $U_m^{(in)}$ and $U_m^{(out)}$, hence
\begin{align*}
\sum_{n=1}^{2\mathcal{N}} \left\{ \int_{u'\in [0, R_1]\cup[R_2, R_3]} f_{n}(u') I(u_m^{(c)},u') u'du' \right\} \sigma_n &= \Phi(u_m^{(c)},0,0) = \begin{cases}
    V_{in}   &u_m^{(c)}\in[0,R_1] \\
    V_{out}   &u_m^{(c)}\in[R_2,R_3]
\end{cases} \\
\sum_{n=1}^{2\mathcal{N}} M_{mn} \sigma_n &= \Phi_m
\end{align*}
with the $M_{mn}$ matrix given by
\[
M_{mn}=\begin{cases}
    \int_{u_{n-1}}^{u_{n}} I(u_m^{(c)},u') u'du' & n=1,\ldots,\mathcal{N} \\
    \int_{u_{n}}^{u_{n+1}} I(u_m^{(c)},u') u'du' & n=N+1,\ldots,2\mathcal{N}. \\
\end{cases} 
\]

\begin{figure}[H]
\centering 
\includegraphics[width=0.4\textwidth]{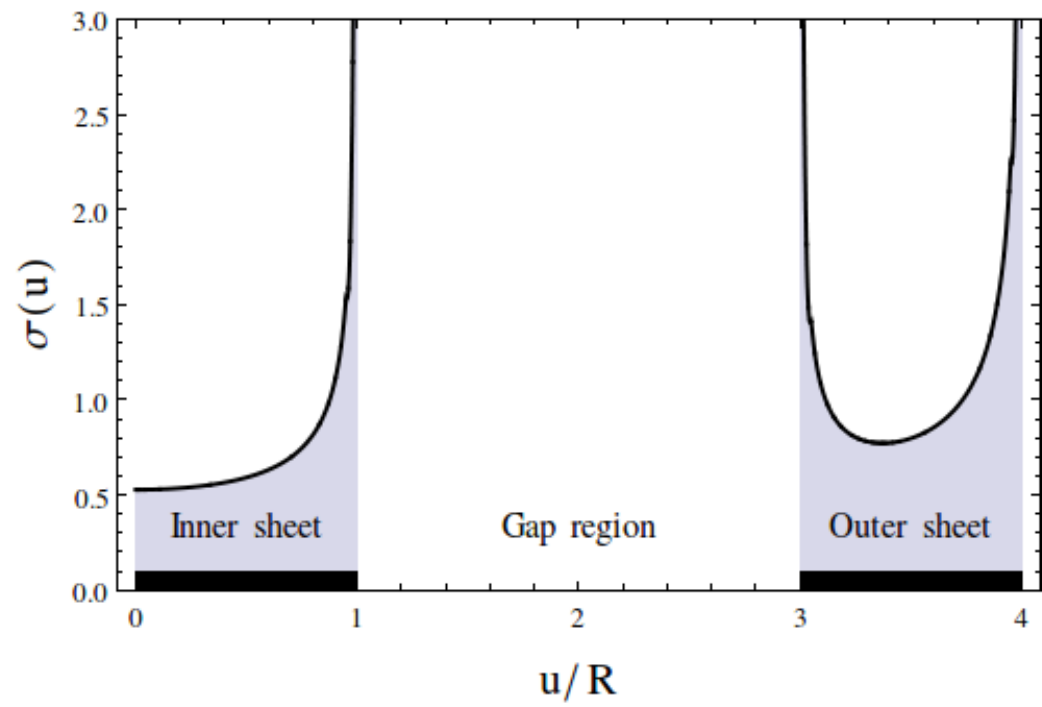}\hspace{1.0cm}
\includegraphics[width=0.39\textwidth]{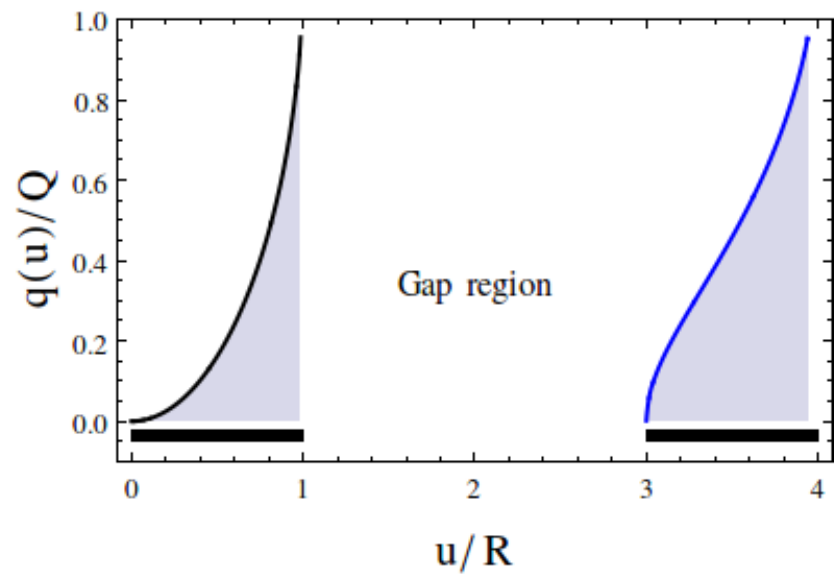}
    \caption[Total charge.]{Illustrative surface density charge (left) and integrated charge profiles (right). Both sheets are at the same voltage $V_{in}=V_{out}=1V$ and their cross sections are represented with black rectangles. }
\label{potentialCurvesEqualFig}
\end{figure} 

Fig.~\ref{potentialCurvesEqualFig}-left shows the profile of the surface charge density $\sigma$ when both sheets are at same potential $V_{in}=V_{out}=1V$ with $R_1=R$, $R_2=3R$ and $R_3=4R$. The potential in the space due to the layers at the same voltage is shown in Fig.~\ref{potentialSurfEqualFig} and the integrated charge $q(u)$ of each layer is shown Fig.~\ref{potentialCurvesEqualIIFig}-left. The last quantity is  computed by 
\begin{equation}
   q(u) = 2\pi\begin{cases}
     \int_{0}^u \sigma(u') du' & \mbox{inner sheet} \\
     \int_{R_2}^u \sigma(u') du' & \mbox{outer sheet}
\end{cases} 
\label{integratedChargeEq}
\end{equation}
with $Q$ the total charge of each layer, this is $Q_{in}=q(R_1)$ and $Q_{out}=q(R_3)$ to normalize the plot. The ratio between total charge is  $Q_{out}/Q_{in} = 9.06972$. The scalar potential in the $\mathbb{R}^3$ space is computed from 
\begin{align*}
\Phi(\boldsymbol{r}) &= \frac{1}{4\pi\epsilon_o} \int_{\Omega_{in}\cup\Omega_{out}} \frac{\sigma(\boldsymbol{r}')}{|\boldsymbol{r}-\boldsymbol{r}'|}d^2\boldsymbol{r} \\   
&=\frac{1}{4\pi\epsilon_o} \int_{u'\in [0, R_1]\cup[R_2, R_3]} \sigma(u') u'du' \int_{0}^{2\pi}  \frac{d\phi'}{ \sqrt{u^2+u'^2 - 2uu'\cos(\phi-\phi') + z^2}} \\
&=\frac{1}{4\pi\epsilon_o} \int_{u'\in [0, R_1]\cup[R_2, R_3]} \sigma(u') u'du' \frac{4}{\sqrt{(u-u')^2+z^2}}K\left(-\frac{4uu'}{(u-u')^2+z^2}\right).
\end{align*}
Here we may use a mid-point approximation, such that the potential takes the form:
\[
\Phi(u,z) = \frac{1}{4\pi\epsilon_o} \sum_{n=1}^{2N} \sigma(u_n^{(c)}) u_n^{(c)} \delta u_n \frac{4}{\sqrt{(u-u_n^{(c)})^2+z^2}}K\left(-\frac{4uu_n^{(c)}}{(u-u_n^{(c)})^2+z^2}\right).
\]


\end{appendices}

\end{document}

%% file: doiCmd.tex
\usepackage{hyperref}

\newcommand{\doi}[1]{{doi:~\href{https://doi.org/#1}{\nolinkurl{#1}}}\rmFullStop}

\newcommand*{\rmFullStop}{\rmifnextchar{.}{}{}}

\makeatletter
\newcommand{\rmifnextchar}[3]{%
  \begingroup
  \ltx@LocToksA{\endgroup#2}%
  \ltx@LocToksB{\endgroup#3}%
  \ltx@ifnextchar{#1}{%
    \def\next{\the\ltx@LocToksA}%
    \afterassignment\next
    \let\scratch= %
  }{%
    \the\ltx@LocToksB
  }%
}
\makeatother

%% file: manuscript.bbl
\begin{thebibliography}{10}

\bibitem{salazar2022electric}
R.~Salazar, C.~Bayona-Roa, and D.~Jaramillo, ``Electric vector potential and
  the biot-savart like law in electrostatics,'' {\em Revista de la Academia
  Colombiana de Ciencias Exactas, F{\'\i}sicas y Naturales}, vol.~46, no.~181,
  pp.~939--946, 2022.
\newblock \doi{10.18257/raccefyn.1671}.

\bibitem{romero2022monte}
R.~Salazar, C.~Bayona, and G.~Tellez, ``Monte carlo simulations of
  two-component coulomb gases applied in surface electrodes,'' {\em Journal of
  Physics: Condensed Matter}, 2022.
\newblock \doi{10.1088/1361-648X/ac4aa8}.

\bibitem{dimonte2008molecular}
G.~Dimonte and J.~Daligault, ``Molecular-dynamics simulations of electron-ion
  temperature relaxation in a classical coulomb plasma,'' {\em Physical review
  letters}, vol.~101, no.~13, p.~135001, 2008.

\bibitem{zelener2018self}
B.~Zelener, B.~Zelener, E.~Manykin, S.~Y. Bronin, A.~Bobrov, and D.~Khikhlukha,
  ``Self-diffusion and conductivity in an ultracold strongly coupled plasma:
  Calculation by the method of molecular dynamics,'' in {\em Journal of
  Physics: Conference Series}, vol.~946, p.~012126, IOP Publishing, 2018.

\bibitem{arkhipov2017direct}
Y.~V. Arkhipov, A.~Askaruly, A.~Davletov, D.~Y. Dubovtsev, Z.~Donk{\'o},
  P.~Hartmann, I.~Korolov, L.~Conde, and I.~Tkachenko, ``Direct determination
  of dynamic properties of coulomb and yukawa classical one-component
  plasmas,'' {\em Physical Review Letters}, vol.~119, no.~4, p.~045001, 2017.

\bibitem{durniak2010molecular}
C.~Durniak, D.~Samsonov, N.~P. Oxtoby, J.~F. Ralph, and S.~Zhdanov,
  ``Molecular-dynamics simulations of dynamic phenomena in complex plasmas,''
  {\em IEEE transactions on plasma science}, vol.~38, no.~9, pp.~2412--2417,
  2010.

\bibitem{zwicknagel1996molecular}
G.~Zwicknagel, C.~Toepffer, and P.-G. Reinhard, ``Molecular dynamic simulations
  of ions in electron plasmas at strong coupling,'' {\em Hyperfine
  Interactions}, vol.~99, no.~1, pp.~285--291, 1996.

\bibitem{calisti2011classical}
A.~Calisti and B.~Talin, ``Classical molecular dynamics model for coupled
  two-component plasmas--ionization balance and time considerations,'' {\em
  Contributions to Plasma Physics}, vol.~51, no.~6, pp.~524--528, 2011.

\bibitem{scheiner2019testing}
B.~Scheiner and S.~D. Baalrud, ``Testing thermal conductivity models with
  equilibrium molecular dynamics simulations of the one-component plasma,''
  {\em Physical Review E}, vol.~100, no.~4, p.~043206, 2019.

\bibitem{mithen2012molecular}
J.~Mithen, J.~Daligault, and G.~Gregori, ``Molecular dynamics simulations for
  the shear viscosity of the one-component plasma,'' {\em Contributions to
  Plasma Physics}, vol.~52, no.~1, pp.~58--61, 2012.

\bibitem{CALISTI2009307}
A.~Calisti, S.~Ferri, and B.~Talin, ``Classical molecular dynamics model for
  coupled two component plasmas,'' {\em High Energy Density Physics}, vol.~5,
  no.~4, pp.~307--311, 2009.

\bibitem{gibson2021method}
W.~C. Gibson, {\em The method of moments in electromagnetics}.
\newblock Chapman and Hall/CRC, 2021.
\newblock \doi{10.1201/9780429355509}.

\bibitem{li2004method}
L.-W. Li, M.-S. Yeo, and M.-S. Leong, ``Method of moments analysis of em fields
  in a multilayered spheroid radiated by a thin circular loop antenna,'' {\em
  IEEE transactions on antennas and propagation}, vol.~52, no.~9,
  pp.~2391--2402, 2004.
\newblock \doi{10.1109/TAP.2004.834018}.

\bibitem{li1999method}
L.-W. Li, C.-P. Lim, and M.-S. Leong, ``Method-of-moments analysis of
  electrically large circular-loop antennas: Nonuniform currents,'' {\em IEE
  Proceedings-Microwaves, Antennas and Propagation}, vol.~146, no.~6,
  pp.~416--420, 1999.
\newblock \doi{10.1049/ip-map:19990784}.

\bibitem{harrington1987method}
R.~F. Harrington, ``The method of moments in electromagnetics,'' {\em Journal
  of Electromagnetic waves and Applications}, vol.~1, no.~3, pp.~181--200,
  1987.
\newblock \doi{10.1163/156939387X00018}.

\bibitem{balanis2016antenna}
C.~A. Balanis, {\em Antenna theory: analysis and design}.
\newblock John wiley \& sons, 2016.

\bibitem{wolfram2012version}
M.~Wolfram, ``Version 9.0,'' {\em Champaign, IL}, 2012.

\bibitem{kumar2022introduction}
G.~Kumar, R.~R. Mishra, and A.~Verma, ``Introduction to molecular dynamics
  simulations,'' in {\em Forcefields for Atomistic-Scale Simulations: Materials
  and Applications}, pp.~1--19, Springer, 2022.
\newblock \doi{10.1007/978-981-19-3092-8}.

\bibitem{leimkuhler2015molecular}
B.~Leimkuhler and C.~Matthews, ``Molecular dynamics,'' {\em Interdisciplinary
  applied mathematics}, vol.~39, p.~443, 2015.
\newblock \doi{10.1007/978-3-319-16375-8}.

\bibitem{salazar2020gaped}
R.~Salazar, C.~Bayona, and G.~T\'ellez, ``Electric vector potential formulation
  in electrostatics: Analytical treatment of the gaped surface electrode,''
  {\em Eur. Phys. J. Plus}, vol.~135, no.~878, 2020.
\newblock \doi{https://doi.org/10.1140/epjp/s13360-020-00864-0}.

\bibitem{salazar2019AngularDependentSE}
R.~Salazar, C.~Bayona, and J.~Chaves, ``Electrostatic field of
  angular-dependent surface electrodes,'' {\em Eur. Phys. J. Plus}, vol.~135,
  no.~93, 2019.
\newblock \doi{10.1140/epjp/s13360-019-00090-3}.

\end{thebibliography}
